\RequirePackage{fix-cm}
\documentclass[smallcondensed]{svjour3}     % onecolumn (ditto)
\smartqed  % flush right qed marks, e.g. at end of proof
\usepackage{graphicx}
\usepackage[numbers]{natbib}
\usepackage{colortbl}
\usepackage{enumerate}
\usepackage{xspace}
\usepackage{bm}
\usepackage{algorithmic}
\usepackage{tikz}
\usetikzlibrary{arrows,decorations.pathmorphing,fit,positioning}
\usepackage{bm}
\usepackage{amsmath}
\usepackage{algorithmic,algorithm}
\usepackage{subcaption}
\usepackage{pgfplots}\pgfplotsset{compat=1.9}
\usepackage{multirow}
\usepackage{longtable}

\newcommand{\ie}{\emph{i.e.,}\xspace}
\newcommand{\eg}{\emph{e.g.,}\xspace}

\newcommand{\paratitle}[1]{\vspace{1.5ex}\noindent \textbf{#1}}

\begin{document}

\title{Multi-label Dataless Text Classification with Topic Modeling%\thanks{Grants or other notes
}


\author{Daochen Zha         \and
        Chenliang Li %etc.
}


\institute{Daochen Zha \at
              State Key Lab of Software Engineering, Wuhan University, China 430079 \\
              \email{daochenzha@whu.edu.cn}           %  \\
           \and
           Chenliang Li \at
              State Key Lab of Software Engineering, Wuhan University, China 430079 \\
              \email{cllee@whu.edu.cn}           %  \\
}

\date{Received: date / Accepted: date}

\maketitle

\begin{abstract}
Manually labeling documents is tedious and expensive, but it is essential for training a traditional text classifier. In recent years, a few \textit{dataless text classification} techniques have been proposed to address this problem. However, existing works mainly center on single-label classification problems, that is, each document is restricted to belonging to a single category. In this paper, we propose a novel \textbf{S}eed-guided \textbf{M}ulti-label \textbf{T}opic \textbf{M}odel, named SMTM. With a few seed words relevant to each category, SMTM conducts multi-label classification for a collection of documents without any labeled document. In SMTM, each category is associated with a single category-topic which covers the meaning of the category. To accommodate with multi-labeled documents, we explicitly model the category sparsity in SMTM by using \textit{spike and slab} prior and weak smoothing prior. That is, without using any threshold tuning, SMTM automatically selects the relevant categories for each document. To incorporate the supervision of the seed words, we propose a seed-guided biased GPU (\ie generalized P\' olya urn) sampling procedure to guide the topic inference of SMTM. Experiments on two public datasets show that SMTM achieves better classification accuracy than state-of-the-art alternatives and even outperforms supervised solutions in some scenarios.
\end{abstract}

\newpage

\section{Introduction}
Multi-label text classification is a fundamental task for textual information organization and management. The task assumes that each document is associated with one or more categories. For example, the paper ``statistical topic models for multi-label document classification''~\cite{rubin2012statistical} can be assigned to multiple categories: \emph{topic model}, \emph{document classification} and \emph{machine learning} simultaneously. In the past decade, many researchers have developed approaches dedicated to automatic multi-label text classification, typically in a supervised manner: (1) manually labeling some sample documents, (2) training a model based on these labeled documents, and (3) assigning category labels automatically on unlabeled documents with the trained model. These approaches often require a considerable number of labeled documents to train a high-quality classifier. However, manually building a multi-labeled training set is much more expensive than a single-labeled counterpart because an annotator needs to consider every possible category for each document. The quality of training set is also hard to control since a user may easily miss some categories when annotating a document.

Many research efforts have been made to reduce the labeling cost in multi-label text classification. Active learning, such as~\cite{yang2009effective,li2013active}, iteratively selects the most informative documents from the unlabeled documents for human annotation. Semi-supervised learning, such as~\cite{soleimani2016semi}, trains a model with both labeled and unlabeled documents. These approaches still need a significant number of labeled documents and remain expensive.

Recently a weakly-supervised setting has emerged to be a promising solution, called \emph{dataless} classification. Without any labeled document, dataless setting assumes that there is a small set of \emph{seed words} relevant to each category. Seed words are much easier to obtain because categories are often meaningful. For example, for \emph{topic model} we can easily select some seed words such as ``topic'', ``LDA'', ``Dirichlet''. In dataless setting, users only need to focus on how to use some seed words to precisely describe each category rather than manually labeling a large number of documents. In this sense, dataless text classification techniques save a lot of human efforts. Recent studies on dataless text classification have achieved great success in single-label classification~\cite{chang2008importance,song2014dataless,chen2015dataless,li2016effective,song2016cross}. It was found that dataless classifiers can achieve close or even better classification accuracy than the state-of-the-art supervised learning solutions~\cite{li2016effective}. However, whether dataless setting is applicable for multi-label text classification is unclear from the literature.

This paper explores whether dataless setting can be applied to multi-label text classification. Multi-label dataless text classification is much more challenging than single-label dataless classification. A major difference of multi-labeled dataset is that a document can be associated with one or more categories. In single-label setting, each document can be modeled with the most relevant category~\cite{li2016effective}, whereas there could be an arbitrary number of categories for each document in multi-label setting. This difference also makes it difficult to assign category labels. Classical classifiers rely on training documents to tune a threshold for each category and apply the threshold on the test set for classification~\cite{fan2007study}. However, this strategy is not applicable in dataless setting since training documents are not available. Another challenge for multi-label dataless classification is how to effectively use the seed words. Typically users can only provide a limited number of seed words because users do not know all the relevant words for a category. A good model needs to deal with the documents that contain no seed word.

In this paper, we propose a \textbf{S}eed-guided \textbf{M}ulti-label \textbf{T}opic \textbf{M}odel to conduct multi-label dataless classification, named SMTM. In SMTM, each category is associated with a single category-topic which covers the meaning of the category. Typically, a document only uses a limited number of category-topics, which we call \emph{category sparsity}. We model the \emph{category sparsity} of the documents by using spike and slab prior and weak smoothing prior~\cite{ishwaran2005spike,lin2014dual}. The spike and slab prior allows us to set up a binary variable between a document and a category. This binary variable works as an ``on/off'' switch to decide whether the category is ``selected'' by the document. The binary variables are naturally decided in a probabilistic manner and can successfully model the category sparsity of the documents in dataless setting. To effectively use the seed words, SMTM resorts to word co-occurence information in the corpus. Specifically, we propose a seed-guided biased GPU sampling procedure for topic inference based on the generalized Po\`lya urn (GPU) model~\cite{mahmoud2008polya}. The proposed sampler promotes the relevant category-topics under a document and the relevant words under a category-topic in ``the rich get richer" manner. In this sense, the semantics of the seed words are propagated during the topic inference procedure. The words that co-occur frequently with the seed words become more likely to be generated from the corresponding category-topic. With few seed words, SMTM is able to continually discover more relevant words which are further used to classify documents.

We evaluate our approach on two public datasets. Experimental results show that SMTM significantly outperforms the state-of-the-art dataless baselines and achieves competitive performance with supervised approaches. We also evaluate some variants of SMTM and discuss the results. Overall, the main contributions of this paper are summarized as follows

\begin{itemize}
\item We propose a novel seed-guided multi-label topic model for multi-label dataless text classification, named SMTM. SMTM is devised to explicitly model the category sparsity of multi-labeled documents by using \textit{spike and slab} prior and weak smoothing prior. To the best of our knowledge, This is the first work to classify documents into relevant categories in a \textit{multi-labeled, and dataless} manner.
\item We introduce a simple yet effective seed-guided biased GPU sampler to guide topic learning process of SMTM. With a few seed words provided for each category, the seed-guided biased GPU sampler can identify the relevant words for each category based on higher-order word co-occurrence patterns, leading to promising classification accuracy in multi-label setting.
\item Our extensive experiments on two public datasets show that the proposed SMTM achieves significantly better classification accuracy than the existing dataless alternatives in terms of both Macro-$F_1$ and Macro-$AUC$ metrics. Also, SMTM even achieves better performance than some supervised solutions in some scenarios.
\end{itemize}

The rest of the paper is organized as follows. In section 2, we summarize the related works of this paper. In section 3, we formalize the problem of multi-label dataless text classification. We then present the proposed SMTM model in detail in Section 4. Section 5 shows the experimental settings, results and discussions. Finally, we conclude the paper in Section 6.

\section{Related Work}
Since our work is related to multi-label text classification, dataless text classification and topic modeling, we review the relevant works from these areas in this section.
\subsection{Dataless Text Classification}
Supervised text classifiers are often hindered by the need for a large number of labeled documents. One thread of the solutions is dataless classification, a weakly-supervised setting that only requires some seed words for each category. The earliest works focus on building a pseudo training set based on the given seed words, which we call \emph{classification-based}. \cite{liu2004text} proposed to use seed words to build an initial training set from the unlabeled documents. Then EM algorithm is applied to train a classifier. \cite{ko2004learning} proposed to construct context-clusters based on seed words. Then a Naive Bayes classifier is learned accordingly by using bootstrapping algorithm. One problem of these approaches is the difficulty of controlling the quality of the pseudo training set, which may lead to unpredictable noisy information into the training procedure.

With the development of semantic representations, some researchers proposed to embed categories, which are represented with seed words, and documents to a shared semantic space. Then the classification is conducted by searching the nearest category for each document, which we call \emph{semantic-based}. \cite{chang2008importance} built a dataless classifier by using Explicit Semantic Analysis (ESA)~\cite{gabrilovich2007computing} based on Wikipedia, showing that dataless classifier is competitive with Naive Bayes in binary classification. \cite{song2014dataless} adapted the semantic method to hierarchical classification and evaluated the effectiveness of a few semantic representations in dataless setting. They found that ESA~\cite{gabrilovich2007computing} performs the best in dataless classification. \cite{song2016cross} further adapted the ESA-based method to cross-lingual dataless classification. For multi-labeled datasets, they treated the multi-label classification as independent binary classification problems and labeled the top K relevant documents as positive for each category. However, the ESA-based method is not well-suited for multi-label text classification. A drawback of this strategy is that it does not consider the imbalanced nature of multi-labeled datasets. This simple adaptation has been included in our experiments.

Another line of dataless classification research is built upon probabilistic models, which we call \emph{probabilistic model based}. \cite{druck2008learning} constrained discriminative probabilistic models with seed words by using generalized expectation (GE) criteria. \cite{chen2015dataless} proposed to use probabilistic topic models to conduct dataless classification. \cite{li2016effective} associated each document with a single category-topic as well as some general topics, then used explicit word co-ocurrence to guide topic models to conduct single-label classification. Probabilistic models bring promising results in dataless classification. However, existing approaches all assume that each document is restricted to belonging to one category so that are not well-suited for multi-label dataless classification.

Our work differs from these works in that we aim to conduct multi-label text classification in dataless setting. To the best of our knowledge, this is the first work to classify documents in a multi-labeled and dataless manner. 

\subsection{Multi-label Text Classification and Topic Models}

Topic models have many applications in a broad range. Latent Dirichlet Allocation (LDA)~\cite{Blei2003latent} is a probabilistic topic model that extracts latent topics underlying a document collection in an unsupervised manner. In the past decade, researchers have adapted LDA to single-label supervised learning, such as Supervised Topic Model~\cite{mcauliffe2008supervised}, DiscLDA~\cite{lacoste2009disclda}, and MedLDA~\cite{zhu2009medlda}.

The first topic model designed for multi-label learning is Labeled-LDA (L-LDA)~\cite{ramage2009labeled}. L-LDA makes a one-to-one correspondence for each category and each topic. In topic inference, L-LDA assumes that each document only uses topics that correspond to its associated categories. They further extended L-LDA to partially labeled LDA by allowing one or more topics for each category~\cite{ramage2011partially}.\cite{rubin2012statistical} improved L-LDA by considering two conditions. They developed Prior-LDA by considering relative frequencies of categories, and Dependancy-LDA by considering dependencies between category labels. \cite{soleimani2016semi} proposed a semi-supervised multi-label topic model (MLTM) to model documents in sentence level. MLTM assigns a category label to a document if and only if at least one sentence in the document is attributed to that category label. MLTM is reported to achieve the state-of-the-art classification results in terms of Macro-AUC~\cite{soleimani2016semi}, thus is included in our experiments.

There is another line of research on discriminative approaches for multi-label text classification. There is a large body of research in this line. Since supervised solution is not the concern of this paper, we only include some representative works. The most simple and frequently used strategy is to adapt single-label classification solutions to the multi-label setting, that is, multi-label classification problem is transformed into a few binary classification problems so that the problem can be solved using binary classifiers~\cite{tsoumakas2006multi,tsoumakas2009mining}. Besides these adaptation based techniques, many researchers have proposed to model interdependencies between categories (\ie category correlation)~\cite{read2011classifier,ghamrawi2005collective,li2016conditional,wang2017regularizing}. Recently, some researchers have also adopted neural network techniques for multi-label classification~\cite{belanger2016structured,cisse2016adios,chen2017ensemble,liu2017deep}.

Most existing works of multi-label text classification assume that there is a set of training documents. Since training documents are often expensive, some strategies have been proposed to save human annotating efforts for multi-label classification. Active learning~\cite{yang2009effective,li2013active} and semi-supervised learning~\cite{soleimani2016semi} are two popular strategies. However, they still require labeled documents and the performance of these approaches still degrades significantly when the size of the training set is not large enough. \cite{tao2012unsupervised} designed an unsupervised framework for multi-label text classification based on the structure of Library of Congress Subject Headings (LCSH). The limitation of their approach is that the labels of LCSH are pre-defined so that their approach is not suited for general multi-label classification tasks. 

Different from these works, we aim to conduct multi-label text classification by using some seed words instead of labeled documents. Our appoach works in a weakly supervised manner such that a lot of human efforts can be saved.

\section{Problem Formalization}
In this section, we formalize the problem of \emph{multi-label dataless text classification}. Let $\mathcal{D} = \{m_1, m_2, ..., m_D \}$ be a set of documents where $D$ is the number of documents. The vocabulary is denoted by $\mathcal{W} = \{1, 2, ..., W\}$ where $W$ is the vocabulary size. Each document $m_d \in \mathcal{D}$ is represented as $\{n_1, n_2, ..., n_{N_d}\}$ where $n_w \in \mathcal{W} (w \in \{1,2,...,N_d\})$ and $N_d$ is the number of tokens in document $m_d$. Let $\mathcal{C} = \{1, 2, ..., C\}$ be a set of categories within the corpus where $C$ is the number of categories. For each category $c \in \mathcal{C}$, we associate it with a small set of seed words, conveying the meaning of the former. The goal of multi-label dataless classification is to classify each document in $\mathcal{D}$ into its relevant categories without the use of labeled documents.

\section{Our Approach}
In this section, we present the proposed \textbf{S}eed-guided \textbf{M}ulti-label \textbf{T}opic \textbf{M}odel for multi-label dataless text classification, named SMTM. We first describe the generative process and inference of our model. Then we introduce the proposed seed-guided biased GPU sampler, which enables effective supervision of seed words for multi-label dataless text classification.

\subsection{Generative Process and Inference}
In SMTM, to model the semantic information relevant to each category, we make a one-to-one correspondence between a category and a topic\footnote{Category and category-topic are considered equivalent and exchangable in this work when the context has no ambiguity.}. We call the topics as \textit{category-topics} in SMTM. Following the strategy used in STM~\cite{li2016effective}, we further introduce a common background topic to model the general semantic information underlying the whole collection. That is, we have $C$ category-topics and a single background topic, where $C$ is the number of categories covered by the collection. A key challenge for the efficacy of SMTM is an appropriate mechanism to model one or more categories of each document. Although a text collection could contain a lot of categories, we observe that each document is likely to belong to only a small number of categories (\ie \emph{category sparsity}). This suggests that each document typically has very few dominating category-topics, \ie the categories of the document. Consider a sentence in a document that belongs to \emph{web} and \emph{education} in Delicius dataset: \emph{``I regularly update my blog, podcast, workshop curricula and social bookmarks.''}. This sentence is expected to use a mixture of three topics, \ie \emph{web}, \emph{education} and the background topic, where the two category-topics are expected to cover the highly-related words (\eg ``blog'', ``podcast'', ``curricula'') and the background topic is expected to include other background words (\eg ``regularly'', ``update''). This modeling strategy is similar to L-LDA～\cite{ramage2009labeled}, in which each document is restricted to using topics that correspond to its categories. However, assigned categories are not available in dataless setting. Here, we propose to automatically select the categories for each document in a probabilistic manner (with the supervision of seed words). Specifically, we utilize \textit{spike and slab} prior and \textit{weak smoothing} prior together to enable the sparsity and the smoothness of the category-topic distribution for a document in SMTM.

The \emph{spike and slab} prior is a well known method to realize ``on'' and ``off'' switch selector in probabilistic topic models~\cite{ishwaran2005spike,wang2009decoupling}. It has been used to learn focused terms or focused topics for better topic inference~\cite{wang2009decoupling,lin2014dual,wang2016targeted}. In SMTM, \emph{spike and slab} prior allows us to associate each document with a set of auxiliary topic selectors which determine whether each corresponding category-topic appears or not. In other words, these topic selectors indicate whether each corresponding category is ``selected'' by a document. Specifically, we introduce an auxiliary Bernoulli variable $\alpha_{d,c}$ to control the presence of category-topic $c$ in document $m_d$. Inspired by \cite{lin2014dual}, we use a regular smoothing prior $\gamma_0$ and a weak smoothing prior $\gamma_1$ ($\gamma_0 \gg \gamma_1$) to decouple the sparsity and smoothness, such that the prior of category-topic distribution for document $d$ is defined to be $\gamma_0\vec{\alpha}_d + \gamma_1\vec{1}$, where $\vec{\alpha}_d=\{\alpha_{d,c}\}_{c=1}^C$ and $\vec{1}$ is the vector with all elements being $1$. The graphical representation of SMTM is shown in Fig.~\ref{fig:plate} and the generative process is described as follows:

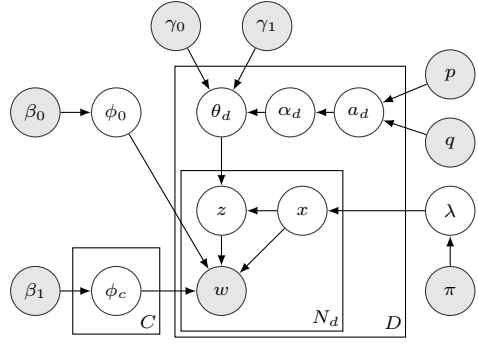
\begin{figure}
\centering
\scriptsize
\begin{tikzpicture}
\tikzstyle{main}=[circle, minimum size = 6.5mm, draw =black!80, node distance = 4mm]
\tikzstyle{connect}=[-latex]
\tikzstyle{box}=[rectangle, draw=black!100, thick]
  \node[main, fill = black!10] (beta_1) {$\beta_1$};
  \node[main] (varphi_c) [right=of beta_1] {$\phi_c$};
  \node[main, node distance = 7.2mm, fill = black!10] (w) [right=of varphi_c] {$w$};
  \node[main, node distance = 4mm] (z) [above=of w] {$z$};
  \node[main, node distance = 4mm] (x) [right=of z] {$x$};
  \node[main, node distance = 12.8mm] (lambda) [right=of x] {$\lambda$};
  \node[main, node distance = 4mm, fill = black!10] (pi) [below=of lambda] {$\pi$};
  \node[main, node distance = 6.4mm] (theta_d) [above=of z] {$\theta_d$};
  \node[main, node distance = 2.4mm] (alpha_d) [right=of theta_d] {$\alpha_d$};
  \node[main, node distance = 2.4mm] (a_d) [right=of alpha_d] {$a_d$};
  \node[main, node distance = 2.4mm, fill = black!10] (q) [above=of lambda] {$q$};
  \node[main, node distance = 2.4mm, fill = black!10] (p) [above=of q] {$p$};
  \node[main, node distance = 4.8mm, fill = black!10] (gamma_1) [above=of theta_d, xshift=6mm] {$\gamma_1$};
  \node[main, node distance = 4.8mm, fill = black!10] (gamma_0) [above=of theta_d,xshift=-6mm] {$\gamma_0$};
  \node[main, node distance = 7.2mm] (varphi_0) [left=of theta_d] {$\phi_0$};
  \node[main, fill = black!10] (beta_0) [left=of varphi_0] {$\beta_0$};
  %\node[main] (theta) [right=of alpha,label=below:$\theta$] { };
  %\node[main] (z) [right=of theta,label=below:z] {};
  %\node[main] (beta) [above=of z,label=below:$\beta$] { };
  %\node[main, fill = black!10] (w) [right=of z,label=below:w] { };
  \path (beta_1) edge [connect] (varphi_c)
        (varphi_c) edge [connect] (w)
    (z) edge [connect] (w)
    (x) edge [connect] (w)
        (varphi_0) edge [connect] (w)
        (x) edge [connect] (z)
        (theta_d) edge [connect] (z)
        (lambda) edge [connect] (x)
        (pi) edge [connect] (lambda)
        (gamma_0) edge [connect] (theta_d)
        (gamma_1) edge [connect] (theta_d)
        (alpha_d) edge [connect] (theta_d)
        (a_d) edge [connect] (alpha_d)
        (p) edge [connect] (a_d)
        (q) edge [connect] (a_d)
        (beta_0) edge [connect] (varphi_0);
  \node[rectangle, inner sep=2.4mm, draw=black!100, fit= (varphi_c)] {};
  \node[rectangle, inner sep=0mm, fit= (varphi_c), label=below right:$C$, xshift=-1.3mm, yshift=1.1mm] {};

  \node[rectangle, inner sep=2mm, draw=black!100, fit= (z) (x) (w)] {};
  \node[rectangle, inner sep=0mm, fit=(w),label=below right:$N_d$, xshift=7.5mm, yshift=1.8mm] {};

  \node[rectangle, inner sep=2.8mm, draw=black!100, fit= (z) (x) (w) (theta_d) (alpha_d) (a_d)] {};
  \node[rectangle, inner sep=0mm, fit=(w),label=below right:$D$, xshift=17mm, yshift=1mm] {};
  %\node[rectangle, inner sep=0mm, fit= (z) (w),label=below right:N, xshift=13mm] {};
 % \node[rectangle, inner sep=4.4mm,draw=black!100, fit= (z) (w)] {};
 % \node[rectangle, inner sep=4.6mm, fit= (z) (w),label=below right:M, xshift=12.5mm] {};
 % \node[rectangle, inner sep=9mm, draw=black!100, fit = (theta) (z) (w)] {};
\end{tikzpicture}
\caption{Graphical representation of SMTM}
\label{fig:plate}
\end{figure}

\begin{enumerate} [1.]
\small
\item Sample a background word distribution $\vec{\phi}_0 \sim Dirichlet(\beta_0)$
\item Sample $\lambda \sim Beta(\pi)$
\item For each category-topic $c \in \{ 1, ..., C\}$:
	\begin{enumerate} [a)]
	\item Sample a word distribution $\vec{\phi}_c \sim Dirichlet(\beta_1)$
	\end{enumerate}
\item For each document $m_d \in \{ m_1, ..., m_D\}$:
	\begin{enumerate} [a)]
	\item Sample $a_{d} \sim Beta(p, q)$
	\item For each category-topic $c \in \{ 1, ..., C\}$:
		\begin{enumerate} [i)]
		\item Sample selector $\alpha_{d,c} \sim Bernoulli(a_{d})$
		\end{enumerate}
    \item Selected category-topic set $A_d = \{k : \alpha_{d,k} = 1\}$
	\item Sample a category-topic distribution $\vec{\theta}_d \sim Dirichlet(\gamma_0 \vec{\alpha}_d + \gamma_1 \vec{1})$
	\item For each position $i \in \{ 1, ..., N_d\}$:
		\begin{enumerate} [i)]
		\item Sample $x_{d,i} \sim Bernoulli(\lambda)$
		\item If $x_{d,i} = 0$, sample $w_{d,i} \sim Multinomial(\vec{\phi}_0)$ \\
		If $x_{d,i} = 1$,
			\begin{enumerate} [1)]
			\item Sample $z_{d,i} \sim Multinomial(\{\theta_{d,k} : k \in A_d \})$
			\item Sample $w_{d,i} \sim Multinomial(\vec{\phi}_{z_{d,i}})$
			\end{enumerate}
		\end{enumerate}
	\end{enumerate}
\end{enumerate}
Here, $z_{d,i}$ is sampled from the selected category-topic set $A_{d}$. Since $\gamma_0 \gg \gamma_1$ ($\gamma_1=10^{-7}$), we can get $\sum_{k \in A_d} \theta_{d,k}=1$ in the numerical sense \cite{lin2014dual}, which results in a much simpler inference procedure.

In SMTM, $\alpha_{d,k}$ is the key to enable multi-label dataless text classification. With the supervision of the seed words (detailed in Section 4.2), the value of $\alpha_{d,k}$ is determined in a probabilistic manner. The switch will be ``on'' (\ie $\alpha_{d,k}=1$) when a document is highly related to a category, or ``off'' (\ie $\alpha_{d,k}=0$) when the document covers less semantic information of the category. That is, the model finds a category set for the document that best fits the document. The sparsity modeling for the document, in turn, enhances the quality of topical words, which further improves the quality of category assignments for other documents. This category selection process continues until a global convergence is reached.

Before introducing how we incorporate the supervision of seed words, we first introduce the inference algorithm for the above described model. We utilize Gibbs sampling for the approximate inference and parameter learning~\cite{griffiths2004finding}. Since $x_{d,i}$ and $z_{d,i}$ are correlated in SMTM, we jointly sample $x_{d,i}$ and $z_{d,i}$ as follows:
\begin{equation}
\begin{aligned}
& P(z_{d,i}, x_{d,i} | \bm{w}, \bm{z}_{\neg di}, \bm{x}_{\neg di}, \bm{\alpha}, \beta_0, \beta_1, \gamma_0, \gamma_1, \pi) \propto \\
&
\begin{cases}
\frac{n^{\neg di}_0 + \pi}{n^{\neg di}_0 + n^{\neg di}_1 + 2 \pi} \times
\frac{n^{\neg di}_{0,w} + \beta_0}{\sum^W_{w'=1}(n^{\neg di}_{0,w'} + \beta_0)}
 & x_{d,i}=0\\
\frac{n^{\neg di}_1 + \pi}{n^{\neg di}_0 + n^{\neg di}_1 + 2\pi}  \times
\frac{n^{\neg di}_{c,w} + \beta_1}{\sum^W_{w'=1}(n^{\neg di}_{c,w'} + \beta_1)} \\ \times
\frac{\alpha_{d,c} n^{\neg di}_{d,c} + \alpha_{d,c} \gamma_0 + \gamma_1}{\sum^C_{c'=1}(\alpha_{d,c'} n^{\neg di}_{d,c'} + \alpha_{d,c'} \gamma_0 + \gamma_1)}

& z_{d,i}=c, x_{d,i}=1
\end{cases}
\end{aligned}
\label{eqn:zx}
\end{equation}
where $n_0$ is the number of words assigned to the background topic, $n_1$ is the number of words assigned to the category-topics, $n_{0,w}$ is the number of times word $w$ is assigned to background topic, $n_{c,w}$ is the number of times word $w$ is assigned to category-topic $c$, $n_{d,c}$ is the number of words assigned to category-topic $c$ in document $d$, $W$ is the vocabulary size, symbol $\neg di$ means that the current assignment is excluded from the count. After sampling all $x_{d,i}$ and $z_{d,i}$, we then sample each category-topic selector $\alpha_{d,c}$ as follows:

\begin{equation}
\begin{aligned}
&P(\alpha_{d,c} | \bm{w}, \bm{z}, \bm{x}, \bm{\alpha}_{\neg dc}, \beta_0, \beta_1, \gamma_0, \gamma_1, \pi) \propto \\
& \begin{cases}
\Gamma(n_{d,c} + \gamma_0 + \gamma_1) \times \Gamma(|\alpha^{\neg c}_d|\gamma_0 + C \gamma_1 + n^{\neg c}_{d,\cdot})  \\\times \Gamma(|\alpha^{\neg c}_d| \gamma_0 + \gamma_0 + C \gamma_1) \times (p + |\alpha^{\neg c}_d|)
 & \alpha_{d,c}=1\\
\Gamma(\gamma_0 + \gamma_1) \times \Gamma(|\alpha^{\neg c}_d|\gamma_0 + \gamma_0 + C \gamma_1 + n^{\neg c}_{d,\cdot})  \\\times \Gamma(|\alpha^{\neg c}_d| \gamma_0 + C \gamma_1) \times (q + C - |\alpha^{\neg c}_d| -1)
 & \alpha_{d,c}=0
 \end{cases}
 \end{aligned}
 \label{eqn:alpha}
 \end{equation}
where $\Gamma(\cdot)$ is the standard Gamma function, $n_{d,\cdot}$ is the sum of $n_{d,c}$ over categories, $|\alpha_d|$ is the number of category-topics selected by document $d$, symbol $\neg c$ means that category $c$ is excluded from the count.

\subsection{Seed-guided Biased GPU Sampler}
Without any supervision, SMTM is just an unsupervised probabilistic topic model. It is difficult to learn the relevant categories for the documents in a purely unsupervised manner. Here, we propose a seed-guided Gibbs sampling procedure by incorporating the seed words through the generalized P\' olya urn (GPU) model~\cite{mahmoud2008polya} in topic inference.

\paratitle{Biased GPU Promotion.} The generalized P\' olya urn (GPU) model represents the discrete probability as colored balls in an urn. The probability of seeing a ball in a color is linearly proportional to the number of balls in that color in the urn. Traditional Gibbs samplers for conventional LDA models follow the simple P\' olya urn (SPU) model. In SPU, when a ball in a particular color is drawn, that ball along with a new ball in the same color is put back, often expressed as ``the rich get richer''. If we perform this repeatedly, the distribution of the colored balls in the urn follows a Dirichlet-multinomial distribution. The GPU differs from SPU in that, when a ball in a particular color is drawn, a certain number of balls in each color are put back along with the drawn ball. This idea has been used in topic modeling for encoding word relatedness knowledge~\cite{mimno2011optimizing,chen2013leveraging,chen2014mining,li2016topic}. That is, a new ball in the same color and also a certain number of balls in similar colors are put back along with the drawn ball. Different from these existing works, we propose a \textit{biased GPU sampler} to guide the sparsity-oriented topic inference of SMTM under the supervision of seed words. By analogy with the GPU model, given a ball of category-topic $c$, we can put back more balls of category-topic $c$ into document $d$ if $d$ is expected to be more likely to belong to category $c$. On the contrary, if we find that document $d$ is less relevant to category $c$, a smaller number of balls of category-topic $c$ will be put back instead. In this sense, the categories that are more related to a particular document will ``get more richer'' under this sampling strategy. A similar procedure can also be applied to guide the word distribution learning for the category-topics in SMTM. Now we explain the strategies for promoting the category-topic distribution and the word distribution.

\paratitle{Promotion for the category-topic distribution.} We observe that a document is likely to belong to a particular category if the document contains the seed words of that category. Thus, we bias the category-topic distribution to prefer a particular category when this document has at least one seed word of that category. More formally, let $\mathcal{I}(d,c)$ be an indicator such that $\mathcal{I}(d,c)=1$ when document $d$ contains at least one seed word of category $c$, and $\mathcal{I}(d,c)=0$ otherwise. The amount of promotion by the biased GPU sampler in SMTM is then calculated as follows:
\begin{align}
u_{c,d}&=
\begin{cases}
1 & \mathcal{I}(c,d)=1\\
\mu & \mathcal{I}(c,d)=0\label{eqn:ucd}\\
 \end{cases}\\
 P_{c,d}&= \frac{u_{c,d}}{\sum_{c'}u_{c',d}} \times C\label{eqn:Pcd}
\end{align}

In Eq.~\ref{eqn:ucd}, $\mu$ is a tunable parameter in the range of $[0,1]$ to control the importance of observing seed words in a document. Based on Eq.~\ref{eqn:Pcd}, when $\mu=1$, no supervision from the seed words is utilized for the calculation of category-topic distributions. That is, the SPU model is recovered instead (\ie $P_{c,d}=1$). When $\mu=0$, the categories covered by a document are restricted to the ones whose seed words appear in the document.

\paratitle{Promotion for the word distribution.} Similarly, we can promote the probabilities of the relevant words for each category-topic. Inspired by \cite{li2016effective}, we use explicit word co-occurrences to estimate the word relevance:
\begin{align}
\label{eqn:Pws}
p(w|s)&=\frac{df(w,s)}{df(s)} \\
v(w,c)&=\frac{1}{|\mathcal{S}_c|} \sum_{s \in \mathcal{S}_c} p(w|s)\\
v_n(w,c)&= max(\frac{v(w,c)}{\sum_{c'}v(w,c')}, \varepsilon)
 \end{align}
where $df(s)$ is the document frequency of a seed word $s$, $df(w,s)$ is the number of documents containing both word $w$ and seed word $s$, $\mathcal{S}_c$ is the set of seed words for category $c$, $\varepsilon$ is a small value to avoid zero ($\varepsilon = 0.01$). Based on the above equations, if word $w$ co-occurs very frequently with seed words of category $c$, $v_n(w,c)$ will have a larger value. Similar to Eq.~\ref{eqn:Pcd}, the promotion for word $w$ under category-topic $c$ is calculated as follows:
\begin{equation}
\tilde{P}_{w,c} = \frac{v_n(w,c)}{\sum_{w'}v_n(w',c)} \times W\label{eqn:Pwc}
\end{equation}

\paratitle{Model inference.} By using GPU model, the joint probability of the words in any topic is not invariant to the permutations of those words. The exact inference, therefore, becomes intractable. Following the work in~\cite{mimno2011optimizing}, we treat each word as if it were the last word, leading to a sampling procedure similar to standard Gibbs sampling based on Eq.~\ref{eqn:zx} and~\ref{eqn:alpha}. The details of the biased GPU sampling process of SMTM is described in Algorithm~\ref{algo:sampling}. When updating $x_{d,i}$ and $z_{d,i}$, $n_{d,z_{d,i}}$ and $n_{z_{d,i},w_{d,i}}$ are sampled based on $P_{z_{d,i},d}$ and $\tilde{P}_{w_{d,i},z_{d,i}}$ respectively. a larger $P_{z_{d,i},d}$ or $\tilde{P}_{w_{d,i},z_{d,i}}$ will encourage $z_{d,i}$ to be sampled from the corresponding category-topic. With the two sampling strategies, the topic inference procedure is effectively supervised by the semantics of the seed words.

\begin{algorithm}[t]
\caption{One iteration of sampling for SMTM}
\label{algo:sampling}
\begin{algorithmic}[1]
\STATE /* update $x_{d,i}$ and $z_{d,i}$ */
\FOR{$d \in \{1, 2, ..., D\}$}
  \FOR{$i \in \{1, 2, ..., N_d\}$}
    \IF{$x_{d,i}=0$}
      \STATE $n_0 \leftarrow n_0 -1$
      \STATE $n_{0,w_{d,i}} \leftarrow n_{0,w_{d,i}}-1$
    \ELSE
      \STATE $n_1 \leftarrow n_1 -1$
      \STATE $n_{d,z_{d,i}} \leftarrow n_{d,z_{d,i}} - P_{z_{d,i},d}$ /* See Eq.~\ref{eqn:Pcd} */
      \STATE $n_{z_{d,i},w_{d,i}} \leftarrow n_{z_{d,i},w_{d,i}} - \tilde{P}_{w_{d,i},z_{d,i}}$ /* See Eq. \ref{eqn:Pwc} */
    \ENDIF
    \STATE sample $x_{d,i}$ and $z_{d,i}$  /* See Eq.~\ref{eqn:zx} */
    \IF{$x_{d,i}=0$}
      \STATE $n_0 \leftarrow n_0 +1$
      \STATE $n_{0,w_{d,i}} \leftarrow n_{0,w_{d,i}}+1$
    \ELSE
      \STATE $n_1 \leftarrow n_1 +1$
      \STATE $n_{d,z_{d,i}} \leftarrow n_{d,z_{d,i}} + P_{z_{d,i},d}$    /* See Eq.~\ref{eqn:Pcd} */
      \STATE $n_{z_{d,i},w_{d,i}} \leftarrow n_{z_{d,i},w_{d,i}} + \tilde{P}_{w_{d,i},z_{d,i}}$  /* See Eq.~\ref{eqn:Pwc} */
    \ENDIF
  \ENDFOR
\ENDFOR
\STATE /* update $\alpha_{d,c}$ */
\FOR{$d \in \{1, 2, ..., D\}$}
  \FOR{$c \in \{1, 2, ..., C\}$}
    \STATE  $|\alpha_{d}| \leftarrow |\alpha_{d}| - \alpha_{d,c}$
    \STATE sample $\alpha_{d,c}$  /* See Eq.~\ref{eqn:alpha} */
    \STATE  $|\alpha_{d}| \leftarrow |\alpha_{d}| + \alpha_{d,c}$
  \ENDFOR
\ENDFOR
\end{algorithmic}
\end{algorithm}

\subsection{Multi-label Classification}
As mentioned above, SMTM automatically selects the relevant categories for each document in a probabilistic manner. For multi-label classification, we assign category label $c$ to document $d$ when the corresponding category-topic is ``selected'' by the document (\ie $\alpha_{d,c} = 1$). For the purpose of performance evaluation, we can also derive a ranking of documents for each category in the descending order of $p(c|d)$. We follow the work in~\cite{li2016topic} to estimate $p(c|d)$ indirectly by using the summation over words (SW) strategy:
 \begin{equation}
\small
p(c|d) \propto \frac{1}{N_d} \sum^{N_d}_{i=1} p(c|w_{d,i})
 \end{equation}
 where $p(c|w_{d,i})$ can be obtained by using Bayes' theorem.

\section{Experiment}
In this section, we conduct extensive experiments to evaluate the performance of SMTM\footnote{We will make the implementation of SMTM publicly available after paper acceptance} on two real-world multi-label datasets. We show that SMTM outperforms existing dataless alternatives. We further examine the scenarios in which SMTM achieves comparable or even better classification performance than the supervised learning solutions. At last, we investigate the impact of different parameter settings and convergence rate, as well as qualitative case study.
\subsection{Datasets}
Two public multi-labeled datasets are used for performance comparison. The first dataset\footnote{http://disi.unitn.it/moschitti/corpora.htm} (called \emph{Ohsumed}) contains medical abstracts from MEDLINE database. Following~\cite{joachims1998text}, we consider the $13,929$ unique abstracts in the first $20,000$ abstracts. The task is to classify the documents into $23$ cardiovascular diseases categories. There are $6,286$ training documents and $7,643$ documents for testing. Each document is associated with $1.7$ categories on average. In our experiments, we use the standard training/test split. The second dataset\footnote{http://nlp.uned.es/social-tagging/delicioust140/} (called \emph{Delicious}) consists of tagged web pages retrieved from social bookmarking service \textit{delicious}~\cite{zubiaga2009content}. Following the work in~\cite{ramage2011partially}, we use the $20$ most common tags for evaluation. For each document, we consider the tags annotated by at least $5$ users. There are $21,670$ documents and each document is associated with $2$ categories on average. We conduct $4$-fold cross validation for this dataset. Both datasets were used previously in evaluating MLTM~\cite{soleimani2016semi}. The datasets are tokenized with NLTK. The stop words, the words shorter than 3 characters and the words appearing in fewer than 5 documents are removed from both datasets. Table \ref{tbl:stat} summarizes the statistics of the two datasets after preprocessing.

\begin{table}
\renewcommand\arraystretch{1}
\caption{Statistics of the two datasets. \#categories: the total number of categories in the dataset; \#documents: the total number of documents in the dataset; \#vocabulary: the size of the vocabulary; \#avgLen: the average number of tokens for each document; \#cardinality: the average number of categories of a document.}
\label{tbl:stat}
\centering
\footnotesize
\begin{tabular}{c|c|c|c|c|c}
\hline
dataset & \#categories & \#documents & \#vocabulary & \#avgLen & \#cardinality\\
\hline
Ohsumed & 23 & 12,929 & 12,711 & 96.67 & 1.66\\
Delicious & 20 & 21,670 & 33,769 & 140.97 & 1.96\\
\hline
\end{tabular}
\end{table}

Following the seed word selection process used in~\cite{chen2015dataless}, we manually select the seed words for each category based on the topical words derived by standard LDA model:

\begin{enumerate}
\item Run standard LDA on the collection to infer latent topics.
\item Manually assign category labels to each topic. If a topic appears not related to any category, we do not assign a category label on that topic.
\item Manually choose at most $10$ seed words based on the most probable $50$ words for each labeled topic.
\end{enumerate}
Note that standard LDA is an unsupervised model that effectively clusters semantically related words, which helps us conduct human selection. There are other approaches that could be used for seed words extraction. (\eg synonymous words or external dictionary). However, the procedure applied here uses no external resource, but merely needs a minimal amount of manual filtering~\cite{chen2015dataless}. After the seed words selection, on average, each category contains $4.3$ and $4.1$ seed words for Ohsumed and Delicious respectively. And a category has a maximum of $7$ seed words for both datasets. These seed words are included in Appendix A.

\subsection{Metrics}
For performance evaluation, we utilize macro-averaged $F_1$ (Macro-$F_1$) and Macro-\textit{AUC}~\cite{ji2008extracting,rubin2012statistical,soleimani2016semi}. Macro-$F_1$ and Macro-\textit{AUC} are the averaged $F_1$ and \textit{AUC} (\ie area under ROC curve) scores of all categories respectively. Let $TP_c$, $FP_c$, $FN_c$ be the the number of true positive, false positive and false negative respectively for category $c$. Then we can obtain Macro-$F_1$ metric as follows:

\begin{equation*}
Macro-F_1 = \frac{1}{C} \sum^{C}_{c=1} \frac{2 \times TP_c}{2 \times TP_c + FP_c + FN_c}
\end{equation*}

Macro-\textit{AUC} is a \emph{ranking-based} metric, \ie it tests the ranking prediction of the most relevant documents for each category. For each category $c$, we first derive a correlation score $S(c, d)$ for each document $d$ based on the classifier. Then we obtain a receiver operating characteristic (ROC) curve by plotting the true positive rate against false positive rate with various threshold settings. The area under the ROC curve is computed for each category and Macro-\textit{AUC} is the average area across categories. A random decision rule gives Macro-\textit{AUC} $= 0.5$; a perfect prediction achieves Macro-\textit{AUC} $= 1$.
\subsection{Baselines}
For thorough comparison, we evaluate SMTM against state-of-the-art supervised and dataless classifiers, as well as some variants of SMTM. We first consider three supervised baselines:
\begin{itemize}
\item \textbf{SVM}: a widely used supervised text classifier. We use a linear kernel and one-vs-rest scheme with TF-IDF weighting. The SVM implementation in sklearn toolkit is used. We tune the penalty parameter $C$ on the set $\{10^i | i=-5,-4, \hdots, 4, 5 \}$.

\item \textbf{L-LDA}: the labeled LDA\footnote{https://nlp.stanford.edu/software/tmt/tmt-0.4/}~\cite{ramage2009labeled} is a topic modeling based supervised approach for multi-label classification. The parameters are tuned and the best result is reported.

\item \textbf{MLTM}: the recently proposed semi-supervised multi-label topic model with sentence-level modeling\footnote{NLTK is used to split the documents into sentences}~\cite{soleimani2016semi}. We use the implementation\footnote{https://github.com/hsoleimani/MLTM} provided by the authors and use their recommended settings. In our experiments, we report the results with full training set, which yields the best results for this model.
\end{itemize}

Tuning decision threshold based on the training set is an important step for supervised multi-label text classification~\cite{fan2007study,rubin2012statistical}. For the supervised classifiers (\ie SVM, L-LDA and MLTM), the threshold is selected using a 4-fold cross-validation. As mentioned in the related work, there are three types of dataless classification techniquess, \ie \emph{classification-based}, \emph{semantic-based}, \emph{probabilistic model based}. For thorough comparison, we consider the state-of-the-art solutions from each type:

\begin{itemize}
\item \textbf{SVM$^s$}, \textbf{L-LDA$^s$} and \textbf{MLTM$^s$}: the straightforward \emph{classification-based} approaches. We build a pseudo training set by associating a training document with a category if the document contains at least one seed word of that category. Then we train a supervised classifier accordingly. We build three dataless classifiers in this setting by using SVM, L-LDA and MLTM respectively. The threshold selection process is the same to supervised selection strategy but conducted over the pseudo training set.

\item \textbf{ESA}: the Explicit Semantic Analysis based dataless classification using Wikipedia~\cite{chang2008importance}, which is the state-of-the-art \emph{semantic-based} solution. The recommended setting is used in our evaluation. We use bootstrapping described in~\cite{song2016cross} since bootstrapping is reported to improve the results of ESA-based methods. We treat the problem as independent binary classification problems and follow~\cite{song2016cross} by labeling the top $K$ relevant documents as positive for each category, where $K$ is tuned and the best result is reported.

\item \textbf{WMD}: word mover's distance is a state-of-the-art word embedding based metric for measuring document distances~\cite{kusner2015word}, which is also a \emph{semantic-based} solution. Similar to ESA, we conduct classification based on semantic distances between documents and categories which are represented with seed words. We implement WMD with the pre-trained 300-dimensional word embeddings\footnote{https://code.google.com/archive/p/word2vec/} from Google News based on Word2Vec~\cite{mikolov2013distributed}.

\item \textbf{DescLDA}: descriptive LDA is a state-of-the-art \emph{probabilistic model based} dataless classifier. Similar to semantic-based solutions, we label the top $K$ relevant documents as positive for each category.
\end{itemize}

Note that STM~\cite{li2016effective} explicitly models only one category for each document so that it is not directly applicable to multi-label classification. We further consider some variants of the proposed model by removing or replacing some components in SMTM:

\begin{itemize}
\item \textbf{SMTM $-$ sparsity}: recall that we explicitly model the category-sparsity in SMTM. Here, we remove the sparsity part in our model. That is, we do not use binary selectors $\vec{\alpha}$. Then we adopt the same top $K$ strategy used in the baselines to conduct multi-label classification.

\item \textbf{SMTM $-$ category promotion}: a variant without the promotion for category-topic distribution. We simply set $\mu=1$ (see Eq.~\ref{eqn:ucd}), so that no supervision from the seed words is utilized for the calculation of category-topic distribution for a document.

\item \textbf{SMTM $-$ word promotion}: a variant without the promotion for word distribution. We fix $\tilde{P}_{w,c} = 1$ (see Eq.~\ref{eqn:Pwc}) so that no bias is incorporated in the sampling process of the words under each category-topic.

\item \textbf{SMTM $+$ word embedding}: recall that SMTM uses explicit word co-occurrences in the target corpus to estimate $P(w|s)$ (See Eq. \ref{eqn:Pws}). We further consider a variant of SMTM that estimates $P(w|s)$ from word embeddings learned from a large external corpus, \ie the pre-trained 300-dimensional word embeddings from Google News based on Word2Vec. Formally, let $cos(s, w)$ denote the cosine similarity between the vector representations of seed word $s$ and word $w$, where $cos(s, w) \in [-1,1]$. Eq.~\ref{eqn:Pws} is rewritten as $P(w|s) = \frac{cos(s, w) + 1}{2}$ such that $P(w|s) \in [0,1]$.

\end{itemize}

For SMTM, we set $\mu=0.3$, $\pi=1$, $p=q=1$, $\beta_0=\beta_1=0.01$, $\gamma_0=50/C$ and $\gamma_1=10^{-7}$. Fast convergence of SMTM is observed in our experiments. We conduct the classification after running SMTM for $100$ iterations. The averaged result over $10$ runs is reported. For fair comparison, the same seed words are used for all dataless methods.

\begin{table}
\footnotesize
\center
\caption{Performance comparison on the two datasets. The best and the second best results by dataless classifiers are highlighted in boldface and underlined respectively. $\#F_1$: Macro-$F_1$ score; $\#AUC$: Macro-$AUC$ score.}

\label{tbl:comparison}
\begin{tabular}{c|c|c||c|c}
\hline
\cline{1-5} & \multicolumn{2}{c||}{Ohsumed} & \multicolumn{2}{c}{Delicious}\\
\cline{2-5}\raisebox{1.5ex}[0pt]{Method} & $\#F_1$ & $\#AUC$ & $\#F_1$ & $\#AUC$\\
\hline
\hline
SVM & 0.629 & 0.921 & 0.461 & 0.846\\
L-LDA & 0.520 & 0.861 & 0.401 & 0.763\\
MLTM & 0.463 & 0.874 & 0.286 & 0.780\\
\hline
SVM$^s$ & 0.418 & 0.789 & 0.340 & 0.754\\
L-LDA$^s$ & 0.411 & 0.818 & 0.321 & 0.745\\
MLTM$^s$ & 0.278 & 0.805 & 0.296 & 0.781\\
ESA & 0.424 & 0.851 & 0.343 & 0.775\\
WMD & 0.264 & 0.753 & 0.268 & 0.783\\
DescLDA & 0.358 & 0.781 & 0.297 & 0.743\\
SMTM & \textbf{0.480} & \textbf{0.872} & \textbf{0.370} & \textbf{0.793}\\
SMTM - sparsity & 0.437 & 0.864 & 0.346 & 0.788\\
SMTM - category promotion & 0.448 & \underline{0.866} & 0.334 & 0.786\\
SMTM - word promotion & 0.450 & 0.861 & 0.362 & \underline{0.789}\\
SMTM + word embedding & \underline{0.451} & 0.845 & \underline{0.364} & 0.783\\
\hline
\end{tabular}
\end{table}

\subsection{Results and Discussion}
The classification performance over the two datasets is reported in Table~\ref{tbl:comparison}. We observe that SMTM significantly outperforms all other dataless methods in terms of both Macro-$F_1$ and Macro-\textit{AUC} on both datasets. Among all dataless baselines in comparison, ESA delivers the best Macro-$F_1$ scores on the two datasets, however with an expensive external knowledge base. We can also find that our approach is much better than DescLDA. Note that DescLDA is also built upon probabilistic topic models but designed for single-label classification. This suggests that our approach successfully discovers the underlying topical structure of multi-labeled documents, leading to better classification results.

As expected, the supervised classifiers like SVM and L-LDA obtain better classification performance than all the dataless classifiers. However, we observe that our approach does the best to close the gap. In fact, the gap between SMTM and L-LDA is small in terms of Macro-$F_1$, and SMTM even achieves better Macro-\textit{AUC} scores than L-LDA. The results of MLTM on both datasets are comparable with the ones reported in \cite{soleimani2016semi}. An interesting finding is that SMTM outperforms MLTM on both datasets in terms of Macro-$F_1$. Note that SMTM and MLTM are both probabilistic topic model based techniques. The superiority of SMTM over MLTM confirms again that explicitly modeling category sparsity of the documents is an effective mechanism for multi-label dataless classification.

SMTM is superior to all its variants. This suggests that the proposed sparsity modeling and promotion strategies are effective. The performance loss when removing either promotion strategy (\ie promotion for the category-topic distribution or the word distribution) validates that the two promotion strategies are complementary so that their combination leads to better classification accuracy. Interestingly, SMTM $+$ word embedding, though with an external resource, underperforms SMTM in all settings. This suggests that using the semantic knowledge in target collection could be more effective than resorting to an external resource, possibly due to the discrepancy in domains.

\subsection{Comparison of SMTM and Supervised Classifiers}
Though SMTM is significantly superior to dataless baselines, from Table~\ref{tbl:comparison}, we observe that SVM is significantly better than our model. Thus, supervised classifiers should be preferred when training data is large in volume and of high-quality. Here, we are interested in conducting a deeper comparison of SMTM and supervised classifiers. Specifically, we will discuss a few scenarios in which our dataless classifier will be a more desired choice than supervised classifiers. 

First, to better understand the performance of our model, we visualize the $F_1$ per category in Fig.~\ref{fig:perlabel}. We observe that, surprisingly, SMTM outperforms L-LDA in terms of $F_1$ for about one third of the categories on both datasets. For some categories SMTM also surpasses SVM. The categories for which SMTM outperforms SVM by more than $3\%$ are \emph{Neoplasms} in the Ohsumed dataset, and \emph{java}, \emph{education} in the Delicious dataset. One possible reason is that these categories are better described by the seed words than the labeled documents. This suggests that seed words could be very strong indicators for some categories, which are even better than labeled documents. That is, seed words could possibly substitute the labeled documents for some categories and result in an even better text classifier. Thus, our approach could possibly be used only to a part of the categories that are precisely described by seed words so that labeled documents are not required for those categories. In this sense, human efforts can be saved with only slight or even no performance loss.

\definecolor{Bluea}{RGB}{201,223,240}
\definecolor{White}{RGB}{253,253,253}
\definecolor{Blueb}{RGB}{235,242,250}
\definecolor{Linecolor}{RGB}{224,234,235}
\definecolor{myblue}{RGB}{59,85,162}
\definecolor{myred}{RGB}{235,125,60}
\definecolor{mypurple}{RGB}{153,50,208}

\begin{figure}
\centering
%\begin{minipage}[t][10cm][t]{.5\textwidth}
%   \vspace*{\fill}
    \centering
\tiny
\centering
    \begin{minipage}{.5\textwidth}
        \centering
        \begin{tikzpicture}
\pgfplotsset{every axis legend/.style={%
cells={anchor=west},
inner xsep=2pt,inner ysep=1pt,nodes={inner sep=1pt,text depth=1pt},
anchor=north east,
shape=rectangle,
draw=black,
at={(0.98,0.28)}
}}
\pgfplotsset{every axis y label/.append style={yshift=0.4cm}}
%\pgfplotsset{compat=1.5}
\begin{axis}[
    height=4.5cm, width=6.5cm,
    xlabel=Category, ylabel=$F_1$,
    every axis x label/.style={
    at={(0.5, -0.1)},
    anchor=north,
},
every axis y label/.style={
    at={(-0.15, 0.55)},
    rotate = 90,
    anchor=east,
},
    grid=both,
    grid style={line width=.1pt, draw=gray!10},
    major grid style={line width=.2pt,draw=gray!50},
    minor x tick num=4,
    minor y tick num=1,
    legend columns=1,
    legend style={fill=white, font=\fontsize{3}{2}\selectfont},
    xtick={0, 5, 10, 15, 20, 25},
    ytick={0, 0.2, 0.4, 0.6, 0.8, 1},
    ymin=0, ymax=1,
    xmin=0, xmax=24,
    ytick pos=left]

    \addplot[
        scatter,only marks,scatter src=explicit symbolic, mark size=2pt, 
        scatter/classes={
            a={mark=*,draw=myblue,fill=myblue},
            b={mark=o,draw=mypurple},
            c={mark=triangle*,myred}
        }
    ]
    table[x=x,y=y,meta=label]{
                x    y    label
                1 0.291 a
                2 0.465 a
                3 0.503 a
                4 0.508 a
                5 0.538 a
                6 0.556 a
                7 0.587 a
                8 0.607 a
                9 0.610 a
                10 0.615 a
                11 0.621 a
                12 0.637 a
                13 0.659 a
                14 0.660 a
                15 0.680 a
                16 0.682 a
                17 0.703 a
                18 0.711 a
                19 0.744 a
                20 0.745 a
                21 0.752 a
                22 0.772 a
                23 0.798 a
                1 0.158 b
                2 0.407 b
                3 0.371 b
                4 0.365 b
                5 0.400 b
                6 0.331 b
                7 0.434 b
                8 0.515 b
                9 0.499 b
                10 0.503 b
                11 0.473 b
                12 0.580 b
                13 0.549 b
                14 0.582 b
                15 0.510 b
                16 0.624 b
                17 0.583 b
                18 0.566 b
                19 0.756 b
                20 0.617 b
                21 0.692 b
                22 0.667 b
                23 0.735 b
                1 0.137 c
                2 0.357 c
                3 0.482 c
                4 0.341 c
                5 0.244 c
                6 0.441 c
                7 0.442 c
                8 0.428 c
                9 0.395 c
                10 0.495 c
                11 0.420 c
                12 0.519 c
                13 0.558 c
                14 0.628 c
                15 0.444 c
                16 0.570 c
                17 0.554 c
                18 0.410 c
                19 0.782 c
                20 0.635 c
                21 0.616 c
                22 0.454 c
                23 0.718 c
            };

    \legend{SVM (averaged 0.629), L-LDA (averaged 0.520), SMTM (averaged 0.480)}

\end{axis}
\end{tikzpicture}
\subcaption{Ohsumed}
    \end{minipage}%
    \begin{minipage}{0.5\textwidth}
        \centering
        \begin{tikzpicture}
\pgfplotsset{every axis legend/.style={%
cells={anchor=west},
inner xsep=2pt,inner ysep=1pt,nodes={inner sep=1pt,text depth=1pt},
anchor=north east,
shape=rectangle,
draw=black,
at={(0.98,0.28)}
}}
\pgfplotsset{every axis y label/.append style={yshift=0.4cm}}
%\pgfplotsset{compat=1.5}
\begin{axis}[
    height=4.5cm, width=6.5cm,
    xlabel=Category, ylabel=$F_1$,
    every axis x label/.style={
    at={(0.5, -0.1)},
    anchor=north,
},
every axis y label/.style={
    at={(-0.15, 0.55)},
    rotate = 90,
    anchor=east,
},
    grid=both,
    grid style={line width=.1pt, draw=gray!10},
    major grid style={line width=.2pt,draw=gray!50},
    minor x tick num=4,
    minor y tick num=1,
    legend columns=1,
    legend style={fill=white, font=\fontsize{3}{2}\selectfont},
    xtick={0, 5, 10, 15, 20},
    ytick={0, 0.2, 0.4, 0.6, 0.8, 1},
    ymin=0, ymax=1,
    xmin=0, xmax=21,
    ytick pos=left]

    \addplot[
        scatter,only marks,scatter src=explicit symbolic, mark size=2pt, 
        scatter/classes={
            a={mark=*,draw=myblue,fill=myblue},
            b={mark=o,draw=mypurple},
            c={mark=triangle*,myred}
        }
    ]
    table[x=x,y=y,meta=label]{
                x    y    label
                1 0.288 a
                2 0.294 a
                3 0.334 a
                4 0.371 a
                5 0.382 a
                6 0.389 a
                7 0.427 a
                8 0.439 a
                9 0.439 a
                10 0.442 a
                11 0.453 a
                12 0.477 a
                13 0.491 a
                14 0.508 a
                15 0.515 a
                16 0.525 a
                17 0.560 a
                18 0.592 a
                19 0.613 a
                20 0.656 a
                1 0.195 b
                2 0.258 b
                3 0.195 b
                4 0.284 b
                5 0.099 b
                6 0.438 b
                7 0.321 b
                8 0.351 b
                9 0.407 b
                10 0.332 b
                11 0.426 b
                12 0.395 b
                13 0.455 b
                14 0.537 b
                15 0.473 b
                16 0.521 b
                17 0.524 b
                18 0.573 b
                19 0.593 b
                20 0.641 b
                1 0.154 c
                2 0.262 c
                3 0.155 c
                4 0.181 c
                5 0.285 c
                6 0.401 c
                7 0.246 c
                8 0.289 c
                9 0.322 c
                10 0.400 c
                11 0.405 c
                12 0.144 c
                13 0.480 c
                14 0.495 c
                15 0.410 c
                16 0.554 c
                17 0.551 c
                18 0.553 c
                19 0.688 c
                20 0.533 c
                
            };

    \legend{SVM (averaged 0.461), L-LDA (averaged 0.401), SMTM (averaged 0.370)}

\end{axis}
\end{tikzpicture}
\subcaption{Delicious}
    \end{minipage}

\caption{Visualizing per category performance in terms of $F_1$. The categories are numbered in increasing order of $F_1$ scores obtained by SVM. The dataless classifier SMTM outperforms L-LDA for $7/23$ categories on the Ohsumed dataset, and $7/20$ categories on the Delicious dataset. SMTM also occasionally outperforms SVM.}
\label{fig:perlabel}
\end{figure}
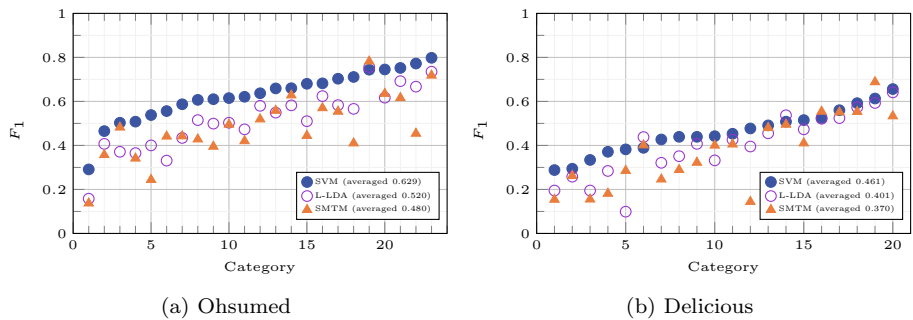

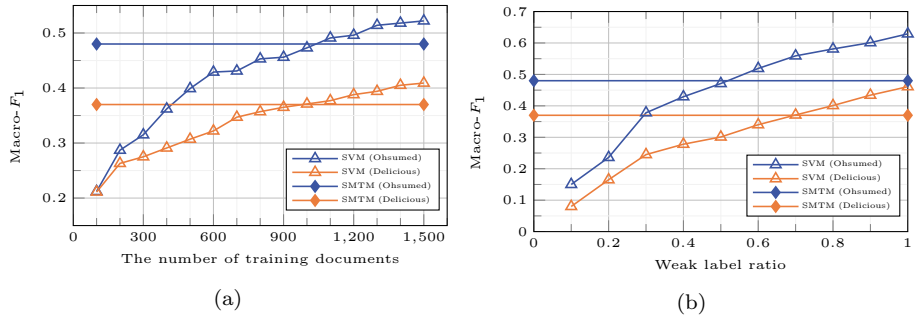
\begin{figure}
\centering
%\begin{minipage}[t][10cm][t]{.5\textwidth}
%   \vspace*{\fill}
    \centering
\tiny
\centering
    \begin{minipage}{.5\textwidth}
        \centering
\begin{tikzpicture}
\pgfplotsset{every axis legend/.style={%
cells={anchor=west},
inner xsep=2pt,inner ysep=1pt,nodes={inner sep=1pt,text depth=1pt},
anchor=north east,
shape=rectangle,
draw=black,
at={(0.98,0.35)}
}}
\pgfplotsset{every axis y label/.append style={yshift=0.4cm}}
%\pgfplotsset{compat=1.5}
\begin{axis}[
    height=4.5cm, width=6.5cm,
    xlabel=The number of training documents, ylabel=Macro-$F_1$,
    every axis x label/.style={
    at={(0.5, -0.1)},
    anchor=north,
},
every axis y label/.style={
    at={(-0.15, 0.66)},
    rotate = 90,
    anchor=east,
},
    grid=both,
    grid style={line width=.1pt, draw=gray!10},
    major grid style={line width=.2pt,draw=gray!50},
    minor x tick num=2,
    minor y tick num=1,
    legend columns=1,
    legend style={fill=white, font=\fontsize{3}{2}\selectfont},
    xtick={0,300,600,900,1200,1500},
    ytick={0.2,0.3,0.4,0.5},
    ymin=0.15, ymax=0.55,
    xmin=0, xmax=1600,
    ytick pos=left]
    \addplot[
        semithick, mark = triangle, mark size=2.3pt,myblue
    ]
    coordinates{
        (100,0.212)
        (200,0.287)
        (300,0.315)
        (400,0.362)
        (500,0.399)
        (600,0.429)
        (700,0.431)
        (800,0.453)
        (900,0.456)
        (1000,0.473)
        (1100,0.491)
        (1200,0.496)
        (1300,0.514)
        (1400,0.518)
        (1500,0.522)
    };

    \addlegendentry{SVM (Ohsumed)}

    \addplot[
        semithick, mark = triangle, mark size=2.3pt,myred
    ]
    %coordinates{(-1,-1)};
     coordinates{
        (100,0.211)
        (200,0.263)
        (300,0.275)
        (400,0.291)
        (500,0.307)
        (600,0.322)
        (700,0.347)
        (800,0.357)
        (900,0.365)
        (1000,0.371)
        (1100,0.377)
        (1200,0.388)
        (1300,0.394)
        (1400,0.405)
        (1500,0.409)
    };
    \addlegendentry{SVM (Delicious)}

    \addplot[
        semithick, mark = diamond*, mark size=2.1pt,myblue
    ]
    coordinates{
        (100,0.480)
        (1500,0.480)
    };
    \addlegendentry{SMTM (Ohsumed)}

    \addplot[
        semithick, mark = diamond*, mark size=2.1pt,myred
    ]
    coordinates{
        (100,0.370)
        (1500,0.370)
    };
    \addlegendentry{SMTM (Delicious)}

\end{axis}
\end{tikzpicture}
\subcaption{}
\label{fig:svm_number}
    \end{minipage}%
    \begin{minipage}{0.5\textwidth}
        \centering
\begin{tikzpicture}
\pgfplotsset{every axis legend/.style={%
cells={anchor=west},
inner xsep=2pt,inner ysep=1pt,nodes={inner sep=1pt,text depth=1pt},
anchor=north east,
shape=rectangle,
draw=black,
at={(0.98,0.35)}
}}
\pgfplotsset{every axis y label/.append style={yshift=0.4cm}}
%\pgfplotsset{compat=1.5}
\begin{axis}[
    height=4.5cm, width=6.5cm,
    xlabel=Weak label ratio, ylabel=Macro-$F_1$,
    every axis x label/.style={
    at={(0.5, -0.1)},
    anchor=north,
},
every axis y label/.style={
    at={(-0.15, 0.66)},
    rotate = 90,
    anchor=east,
},
    grid=both,
    grid style={line width=.1pt, draw=gray!10},
    major grid style={line width=.2pt,draw=gray!50},
    minor x tick num=1,
    minor y tick num=1,
    legend columns=1,
    legend style={fill=white, font=\fontsize{3}{2}\selectfont},
    xtick={0, 0.2, 0.4, 0.6, 0.8, 1},
    ytick={0,0.1,0.2,0.3,0.4,0.5,0.6,0.7},
    ymin=0, ymax=0.7,
    xmin=0, xmax=1,
    ytick pos=left]
    \addplot[
        semithick, mark = triangle, mark size=2.3pt,myblue
    ]
    coordinates{
        (0.1,0.150)
        (0.2,0.236)
        (0.3,0.378)
        (0.4,0.429)
        (0.5,0.471)
        (0.6,0.519)
        (0.7,0.559)
        (0.8,0.581)
        (0.9,0.601)
        (1,0.629)

    };

    \addlegendentry{SVM (Ohsumed)}

    \addplot[
        semithick, mark = triangle, mark size=2.3pt,myred
    ]
    %coordinates{(-1,-1)};
     coordinates{
        (0.1,0.08)
        (0.2,0.165)
        (0.3,0.245)
        (0.4,0.278)
        (0.5,0.301)
        (0.6,0.340)
        (0.7,0.371)
        (0.8,0.401)
        (0.9,0.434)
        (1,0.461)

    };
    \addlegendentry{SVM (Delicious)}

    \addplot[
        semithick, mark = diamond*, mark size=2.1pt,myblue
    ]
    coordinates{
        (0,0.480)
        (1,0.480)
    };
    \addlegendentry{SMTM (Ohsumed)}

    \addplot[
        semithick, mark = diamond*, mark size=2.1pt,myred
    ]
    coordinates{
        (0,0.370)
        (1,0.370)
    };
    \addlegendentry{SMTM (Delicious)}

\end{axis}
\end{tikzpicture}
\subcaption{}
\label{fig:svm_partial}
    \end{minipage}

\caption{The performance comparision of SMTM and SVM. (a) plots the performance of SVM with different number of training documents; (b) plots the performance of SVM with different weak label ratio, that is, only that ratio of category labels are kept for each document.}
\label{fig:svm}
\end{figure}

Second, the volume of training data is critical to the effectiveness of supervised classifiers. Here, we would like to find the number of training instances at which SVM starts to outperform SMTM. Following the work in~\cite{chen2015dataless}, we randomly create the subsets from training documents such that the proportion of documents under each category is identical to that of the whole training set. The performance pattern of SVM in terms of Macro-$F_1$ is shown in Fig.~\ref{fig:svm_number}. It takes about $1,000$ training documents for SVM to obtain comparable performance with SMTM on both datasets. Note that SMTM only uses about 4 seed words for each category on average. SMTM would be a desired choice when a large number of labeled documents are not available.

Third, in many applications, there is only a ``partial" set of category labels for a document in the training set~\cite{sun2010multi}. To prepare a multi-labeled dataset, a user has to consider every possible category for each document. It is likely that the user will miss some proper categories. When a user provides a particular category label for a document, we know that the category is proper. However, for the categories not provided, we can not conclude that they are not proper. Regularly, ``partial" category labels will degrade the performance of supervised classifiers. We randomly use a specific ratio of the category labels for each document and train an SVM classifier accordingly. The result is reported in Fig.~\ref{fig:svm_partial}. We observe that the performance of SMTM is close to SVM using about $50\%$ to $70\%$ of the category labels for each document. For supervised classifiers, it is expensive to ask human annotators to carefully consider all the categories and assign a perfect category label set for each document. Our dataless classifier will be preferable when the quality of labeled documents cannot be guaranteed.

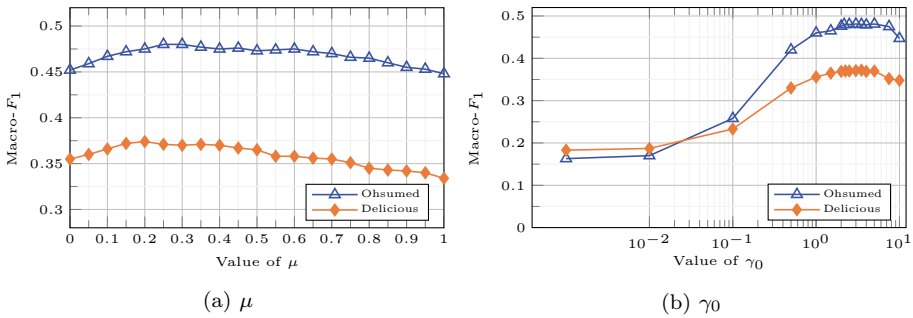
\begin{figure}
\centering
%\begin{minipage}[t][10cm][t]{.5\textwidth}
%   \vspace*{\fill}
    \centering
\tiny
\centering
    \begin{minipage}{.5\textwidth}
        \centering
\begin{tikzpicture}
\pgfplotsset{every axis legend/.style={%
cells={anchor=west},
inner xsep=2pt,inner ysep=1pt,nodes={inner sep=1pt,text depth=1pt},
anchor=north east,
shape=rectangle,
draw=black,
at={(0.96,0.20)}
}}
\pgfplotsset{every axis y label/.append style={yshift=0.4cm}}
\begin{axis}[
    height=4.5cm, width=6.5cm,
    grid=both,
    grid style={line width=.1pt, draw=gray!10},
    major grid style={line width=.2pt,draw=gray!50},
    minor y tick num=1,
    minor x tick num=1,
    xlabel=Value of $\mu$, ylabel=Macro-$F_1$,
    every axis x label/.style={
    at={(0.5, -0.1)},
    anchor=north,
},
every axis y label/.style={
    at={(-0.15, 0.66)},
    rotate = 90,
    anchor=east,
},
    legend columns=1,
    legend style={fill = white, font=\fontsize{4}{5}\selectfont},
    xtick={0, 0.1, 0.2, 0.3, 0.4, 0.5, 0.6, 0.7, 0.8, 0.9, 1.0},
    xmin=0, xmax=1.0,
    ytick={0.30,0.35,0.40,0.45,0.50},
    ymin=0.28, ymax=0.52,
    ytick pos=left]
    \addplot[
        semithick, mark = triangle, mark size=2.3pt,myblue
    ]
    coordinates{
        (0, 0.452)
        (0.05, 0.459)
        (0.10, 0.467)
        (0.15, 0.472)
        (0.20, 0.475)
        (0.25, 0.480)
        (0.30, 0.480)
        (0.35, 0.477)
        (0.40, 0.475)
        (0.45, 0.476)
        (0.50, 0.473)
        (0.55, 0.474)
        (0.60, 0.475)
        (0.65, 0.472)
        (0.70, 0.470)
        (0.75, 0.466)
        (0.80, 0.465)
        (0.85, 0.460)
        (0.90, 0.455)
        (0.95, 0.453)
        (1.00, 0.448)
    };

    \addlegendentry{Ohsumed}

    \addplot[
        semithick, mark = diamond*, mark size=2.1pt,myred
    ]
    coordinates{
        (0, 0.355)
        (0.05, 0.360)
        (0.10, 0.366)
        (0.15, 0.372)
        (0.20, 0.374)
        (0.25, 0.371)
        (0.30, 0.370)
        (0.35, 0.371)
        (0.40, 0.370)
        (0.45, 0.367)
        (0.50, 0.365)
        (0.55, 0.358)
        (0.60, 0.358)
        (0.65, 0.356)
        (0.70, 0.355)
        (0.75, 0.351)
        (0.80, 0.345)
        (0.85, 0.343)
        (0.90, 0.342)
        (0.95, 0.340)
        (1.00, 0.334)
    };
     \addlegendentry{Delicious}
\end{axis}

\end{tikzpicture}
\subcaption{$\mu$}
    \end{minipage}%
    \begin{minipage}{0.5\textwidth}
        \centering
\begin{tikzpicture}
\pgfplotsset{every axis legend/.style={%
cells={anchor=west},
inner xsep=2pt,inner ysep=1pt,nodes={inner sep=1pt,text depth=1pt},
anchor=north east,
shape=rectangle,
draw=black,
at={(0.96,0.20)}
}}
\pgfplotsset{every axis y label/.append style={yshift=0.4cm}}
\begin{axis}[
    xmode=log,
    height=4.5cm, width=6.5cm,
    grid=both,
    grid style={line width=.1pt, draw=gray!10},
    major grid style={line width=.2pt,draw=gray!50},
    minor y tick num=1,
    minor x tick num=0,
    xlabel=Value of $\gamma_0$, ylabel=Macro-$F_1$,
    every axis x label/.style={
    at={(0.5, -0.1)},
    anchor=north,
},
every axis y label/.style={
    at={(-0.15, 0.66)},
    rotate = 90,
    anchor=east,
},
    legend columns=1,
    legend style={fill = white, font=\fontsize{4}{5}\selectfont},
    xtick={0.01, 0.1, 1, 10},
    xmin=0, xmax=12,
    ytick={0,0.1,0.2,0.3,0.4,0.5},
    ymin=0, ymax=0.52,
    ytick pos=left]
    \addplot[
        semithick, mark = triangle, mark size=2.3pt,myblue
    ]
    coordinates{
        (0.001, 0.163)
        (0.01, 0.170)
        (0.1, 0.258)
        (0.5, 0.420)
        (1, 0.460)
        (1.5, 0.465)
        (2, 0.476)
        (2.17, 0.480)
        (2.5, 0.480)
        (3, 0.481)
        (3.5, 0.480)
        (4, 0.479)
        (5, 0.481)
        (7.5, 0.475)
        (10, 0.447)

    };

    \addlegendentry{Ohsumed}

    \addplot[
        semithick, mark = diamond*, mark size=2.1pt,myred
    ]
    coordinates{
        (0.001, 0.183)
        (0.01, 0.187)
        (0.1, 0.233)
        (0.5, 0.330)
        (1, 0.356)
        (1.5, 0.365)
        (2, 0.369)
        (2.25, 0.370)
        (2.5, 0.370)
        (3, 0.371)
        (3.5, 0.372)
        (4, 0.369)
        (5, 0.370)
        (7.5, 0.352)
        (10, 0.348)
    };
     \addlegendentry{Delicious}
\end{axis}

\end{tikzpicture}
\subcaption{$\gamma_0$}
    \end{minipage}

\caption{The impact of parameters in SMTM.}
\label{fig:parameter}
\end{figure}
\begin{figure}
%\begin{minipage}[t][10cm][t]{.5\textwidth}
%   \vspace*{\fill}
    \centering
\tiny
\begin{tikzpicture}
\pgfplotsset{every axis legend/.style={%
cells={anchor=west},
inner xsep=2pt,inner ysep=1pt,nodes={inner sep=1pt,text depth=1pt},
anchor=north east,
shape=rectangle,
draw=black,
at={(0.96,0.20)}
}}
\pgfplotsset{every axis y label/.append style={yshift=0.4cm}}
\begin{axis}[
    height=4.5cm, width=6.5cm,
    grid=both,
    grid style={line width=.1pt, draw=gray!10},
    major grid style={line width=.2pt,draw=gray!50},
    minor y tick num=1,
    minor x tick num=4,
    xlabel=The number of iterations, ylabel=Macro-$F_1$,
    every axis x label/.style={
    at={(0.5, -0.1)},
    anchor=north,
},
every axis y label/.style={
    at={(-0.15, 0.66)},
    rotate = 90,
    anchor=east,
},
    legend columns=1,
    legend style={fill = white, font=\fontsize{4}{5}\selectfont},
    xtick={0,5,10,15,20},
    xmin=1, xmax=20,
    ytick={0.2,0.3,0.4,0.5},
    ymin=0.2, ymax=0.52,
    ytick pos=left]
    \addplot[
        semithick, mark = triangle, mark size=2.3pt,myblue
    ]
    coordinates{
        (1, 0.247)
        (2, 0.319)
        (3, 0.375)
        (4, 0.414)
        (5, 0.434)
        (6, 0.445)
        (7, 0.459)
        (8, 0.463)
        (9, 0.480)
        (10, 0.481)
        (11, 0.479)
        (12, 0.478)
        (13, 0.478)
        (14, 0.480)
        (15, 0.482)
        (16, 0.478)
        (17, 0.480)
        (18, 0.483)
        (19, 0.479)
        (20, 0.478)

    };

    \addlegendentry{Ohsumed}

    \addplot[
        semithick, mark = diamond*, mark size=2.1pt,myred
    ]
    coordinates{
        (1, 0.256)
        (2, 0.294)
        (3, 0.330)
        (4, 0.348)
        (5, 0.366)
        (6, 0.370)
        (7, 0.367)
        (8, 0.368)
        (9, 0.370)
        (10, 0.369)
        (11, 0.371)
        (12, 0.371)
        (13, 0.372)
        (14, 0.373)
        (15, 0.372)
        (16, 0.370)
        (17, 0.370)
        (18, 0.369)
        (19, 0.368)
        (20, 0.371)

    };
     \addlegendentry{Delicious}
\end{axis}

\end{tikzpicture}
%\end{minipage}
\caption{The performance of SMTM with different number of iterations.}
\label{fig:iter}
\vspace{-0.3cm}
\end{figure}
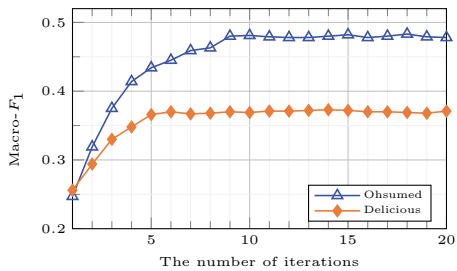

\subsection{Impact of Parameters and the Number of Iterations}
Our model has a few parameters. We study the impact of those parameters by varying each parameter respectively with other parameters fixed. Note that we do not consider $\gamma_1$ and fix it to $10^{-7}$ since the Gibbs sampling algorithm holds only when $\gamma_1$ is very small. In our experiments, we find that our model is not sensitive to most parameters, except $\mu$ and $\gamma_0$. The performance patterns of SMTM with different $\mu$ and $\gamma_0$ values are plotted in Fig.~\ref{fig:parameter}. The two parameters can be well interpreted. $\mu$ controls the importance of observing seed words. Small $\mu$ value indicates that the appearance of a seed word is a very strong indicator of the corresponding category. $\gamma_0$ controls the category sparsity of each document and can be understood as a ``threshold" of assigning a category label. A small $\gamma_0$ value means a document will be assigned more category labels, that is, a category will be included even when this category is only weakly related to a document. A large $\gamma_0$ results in a small number of category labels for each document, which means a category will be included only when this category is highly related to a document. We can observe that a larger $\gamma_0$ (\ie about $1$ to $5$) leads to better performance, consistent with the assumption that the category-topic distribution for a document is sparse. In our evaluation, we set $\gamma_0 = 50 / C$ on both datasets, which is a typical value for topic models. For other insensitive parameters, we also set them to typical values.

Fast convergence of SMTM is observed. The performance with respect to the number of iterations is plotted in Fig.~\ref{fig:iter}. We observe stable performance after about $10$ iterations.

\subsection{Topic Visualization and Case Study}
Table~\ref{tbl:topic} shows the most probable words under some sample category-topics learned from the Delicious dataset by using SMTM and L-LDA. Observe that, in addition to the seed words, SMTM discovers other relevant words for each category. For example, SMTM finds ``obama'' in category \emph{politics}, and ``desktop'' in category \emph{computer}, although both are not seed words. We also observe that some irrelevant words are discovered by L-LDA, \eg ``new'', ``one'', ``use'' and ``would''. Note that L-LDA uses a large number of labeled documents in learning. The above observations show that SMTM can discover meaningful category-topics that are comparable with supervised topic models.

\begin{table*}
\footnotesize
\center
\caption{Top-10 words learned from the Delicious dataset by SMTM (left) and L-LDA (right) for sample categories. The seed words of SMTM are highlighted in boldface, and irrelevant words are underlined.}
\label{tbl:topic}
\begin{tabular}{c!{\color{Linecolor}\vrule}l!{\color{Linecolor}\vrule}l}
\arrayrulecolor{Linecolor}\hline
\rowcolor{Bluea} Category & \multicolumn{1}{c!{\color{Linecolor}\vrule}}{SMTM top-10 words} & \multicolumn{1}{c}{L-LDA top-10 words}\\
\arrayrulecolor{Linecolor}\hline
\rowcolor{White} ~ & ~ & ~\\
\rowcolor{White} ~ & ~ & ~\\
\rowcolor{White} \multirow{-3}{*}{style} & \multirow{-3}{4cm}{\textbf{style} color styles fonts div cascading colors width font sheets}
 & \multirow{-3}{4cm}{style css color \underline{use} page font \underline{name} display styles fonts}\\
 \arrayrulecolor{Linecolor}\hline
\rowcolor{Blueb} ~ & ~ & ~\\
\rowcolor{Blueb} ~ & ~ & ~\\
\rowcolor{Blueb} ~ & ~ & ~\\
\rowcolor{Blueb} \multirow{-4}{*}{politics} & \multirow{-4}{4cm}{\textbf{government} obama \textbf{political} mccain \textbf{politics} presidential campaign \textbf{senate} \textbf{democracy} federal}
 & \multirow{-4}{4cm}{government political \underline{new} people world obama public campaign mccain \underline{would}}\\
 \arrayrulecolor{Linecolor}\hline
\rowcolor{White} ~ & ~ & ~\\
\rowcolor{White} ~ & ~ & ~\\
\rowcolor{White} \multirow{-3}{*}{computer} & \multirow{-3}{4cm}{\textbf{mac} \textbf{computer} desktop \textbf{hardware} screen windows \textbf{drive} apple linux usb}
 & \multirow{-3}{4cm}{mac computer \underline{use} windows software free download file \underline{new} apple}\\
\arrayrulecolor{Linecolor}\hline
\rowcolor{Blueb} ~ & ~ & ~\\
\rowcolor{Blueb} ~ & ~ & ~\\
\rowcolor{Blueb} \multirow{-3}{*}{culture} & \multirow{-3}{4cm}{\textbf{music} \textbf{art} \textbf{culture} artists artist film festival \underline{stock} songs arts}
 & \multirow{-3}{4cm}{art music \underline{new} \underline{one} culture artists time work world video}\\
 \arrayrulecolor{Linecolor}\hline
\end{tabular}
\end{table*}

Table \ref{tbl:case} shows three documents in the Delicious dataset. We make the following observations. First, most of the words are generated from the background topic in SMTM. It is reasonable because regularly only those words highly relevant to a category are expected to be generated from the corresponding category-topic. Second, each document only uses category-topics that correspond to its prediction, which provides evidence of the effect of sparsity modeling of SMTM. Third, SMTM can correctly classify a document even if the document contains no seed word. For example, in the second document, there is no seed word of \emph{politics}. Nevertheless, SMTM successfully associates the document with \emph{politics}. An explanation is that some words in the document (\eg ``democratic", ``obama") frequently co-occur with the seed words of \emph{politics}, and are identified as relevant words by our model. This also explains why SMTM can discover relevant words in addition to seed words, as illustrated in Table \ref{tbl:topic}. Fourth, some documents may be irrelevant to a particular category even though it contains relevant words of that category. For example, the third document mentions ``web" only to tell the readers how to use the tools, but not to talk about the \emph{web}. Unfortunately, ``web" is the seed word of \emph{web} in our experiments. It is not a surprise that SMTM mistakenly associates the document with \emph{web} since seed words are assumed to be strong indicators of corresponding categories. Our model could be further improved by considering these conditions.

\definecolor{caseblue}{RGB}{59,85,162}
\definecolor{casered}{RGB}{238,92,66}
\definecolor{casepurple}{RGB}{153,50,208}
\definecolor{caseorange}{RGB}{228,144,0}
\definecolor{casegreen}{RGB}{83,168,83}
\definecolor{casegrey}{RGB}{204,204,204}

\begin{table}
\footnotesize
\center
\caption{Topic assignments of some documents in the Delicious dataset. The words and punctuations that are removed in the preprocessing step are shaded with gray. Black words are generated from background topic; red from \emph{education}; blue from \emph{web}; purple from \emph{politics}. The seed words are underlined. Note that there is no seed word in the second document.}
\label{tbl:case}
\scriptsize
\begin{tabular}{|l|p{9cm}|}

\hline Ground Truth:& 
\multirow{10}{9cm}{{\color{casegrey}Hi! My} name {\color{casegrey}is} Wesley Fryer{\color{casegrey}. I am} thrilled {\color{casegrey}to be a 21st} century digital {\color{casered}learner}{\color{casegrey}. We} live {\color{casegrey}in the most} exciting age {\color{casegrey}of} earth history {\color{casegrey}for} anyone {\color{casegrey}with} ideas {\color{casegrey}they} want {\color{casegrey}to} share {\color{casegrey}with a} global audience{\color{casegrey}! As I} process {\color{casegrey}the} world {\color{casegrey}and the} {\color{caseblue}\underline{web}}{\color{casegrey}. I} regularly update {\color{casegrey}my} {\color{caseblue}blog}{\color{casegrey},} {\color{caseblue}podcast}{\color{casegrey},} workshop {\color{casered}curricula} {\color{casegrey}and} social bookmarks{\color{casegrey}. (You can} add {\color{casegrey}me to your own del.icio.us} network{\color{casegrey}.) I} frequently contribute {\color{casegrey}to} {\color{casered}Technology} {\color{casegrey}and} {\color{casered}\underline{Learning}}{\color{casegrey}'s} {\color{caseblue}Blog} {\color{casegrey}and} {\color{caseblue}Google}{\color{casegrey}'s} {\color{casered}\underline{Education}} {\color{caseblue}blog}{\color{casegrey}, ``The \color{black}Infinite Thinking Machine\color{casegrey}."} {\color{casegrey}A} fairly complete list {\color{casegrey}of the} websites {\color{casegrey}I} maintain{\color{casegrey},} social networks {\color{casegrey}to which I} regularly contribute{\color{casegrey}, and other} {\color{caseblue}\underline{web}} {\color{casegrey}2.0} sites {\color{casegrey}I} utilize {\color{casegrey}is} available {\color{casegrey}on claimid.com/wfryer.}}
\\
\emph{\color{caseblue} web}, \emph{\color{casered} education}&~\\
Prediction:&~\\
\emph{\color{caseblue} web}, \emph{\color{casered} education}&~\\
~&~\\
~&~\\
~&~\\
~&~\\
~&~\\
~&~\\

\hline Ground Truth:&
\multirow{10}{9cm}{{\color{casegrey}It's} time {\color{casegrey}for} presumed {\color{casepurple}Democratic nominee Barack Obama} {\color{casegrey}to} turn {\color{casegrey}his} attention {\color{casegrey}to a} running mate{\color{casegrey}. To} help{\color{casegrey}, we} bring {\color{casegrey}you the} second installment {\color{casegrey}of VP} Madness{\color{casegrey}, where} users decide {\color{casegrey}who} {\color{casepurple}Obama} {\color{casegrey}should} choose {\color{casegrey}as his \#2.} Vote {\color{casegrey}in the head-to-head} match ups {\color{casegrey}below to} determine {\color{casegrey}which} {\color{casepurple}candidates} advance {\color{casegrey}to} face {\color{casegrey}each other in the} next round{\color{casegrey}. You can} view {\color{casegrey}the} latest results {\color{casegrey}by} clicking {\color{casegrey}the} button {\color{casegrey}at the} bottom {\color{casegrey}of the} page{\color{casegrey}. The} winner {\color{casegrey}will be} revealed on July {\color{casegrey}1, in} plenty {\color{casegrey}of} time {\color{casegrey}for} {\color{casepurple}Obama} {\color{casegrey}to} consider {\color{casegrey}your} choice{\color{casegrey}. In the} {\color{casepurple}GOP} Edition{\color{casegrey},} former {\color{casepurple}Arkansas Governor} {\color{casegrey}(and} former {\color{casepurple}presidential candidate}{\color{casegrey})} Mike {\color{casepurple}Huckabee} was chosen {\color{casegrey}as} John {\color{casepurple}McCain}{\color{casegrey}'s} best bet{\color{casegrey}.}}
\\
\emph{\color{casepurple} politics}&~\\
Prediction:&~\\
\emph{\color{casepurple} politics}&~\\
~&~\\
~&~\\
~&~\\
~&~\\
~&~\\
~&~\\

\hline Ground Truth:&
\multirow{7}{9cm}{{\color{casegrey}A} note {\color{casegrey}about the} resources presented{\color{casegrey}: The} following {\color{casegrey}is a} collection {\color{casegrey}of} audio{\color{casegrey},} video {\color{casegrey}and} multimedia {\color{casered}\underline{learning}} tools {\color{casegrey}for} use {\color{casegrey}by} {\color{casered}faculty} {\color{casegrey}and} {\color{casered}\underline{students}}{\color{casegrey}.} {\color{casegrey}To} use {\color{casegrey}any of the} tools {\color{casegrey}below you can} link {\color{casegrey}to this} page {\color{casegrey}as} needed {\color{casegrey}or} simply {\color{casegrey}right-click your} mouse {\color{casegrey}on the} title{\color{casegrey}; then} copy {\color{casegrey}the} {\color{caseblue}\underline{web}} {\color{caseblue}address} {\color{casegrey}(}{\color{caseblue}shortcut}{\color{casegrey},} {\color{caseblue}URL}{\color{casegrey},} link location{\color{casegrey}) to your} system{\color{casegrey}'s} clipboard{\color{casegrey}; and} {\color{caseblue}paste} {\color{casegrey}the} direct {\color{caseblue}URL} {\color{casegrey}into your} {\color{caseblue}code}{\color{casegrey}. The} use {\color{casegrey}of these} objects {\color{casegrey}is} free {\color{casegrey}for} {\color{casered}nonprofit educational} use {\color{casegrey}with} proper attribution {\color{casegrey}to the CIP as} author{\color{casegrey}.}}
\\

\emph{\color{casered} education}&~\\
Prediction:&~\\
\emph{\color{caseblue} web}, \emph{\color{casered} education}&~\\
~&~\\
~&~\\
~&~\\
\hline
\end{tabular}
\end{table}

\section{Conclusion and Future Works}
In this paper, we proposed a novel Seed-guided Multi-label Topic Model for multi-label dataless text classification, named SMTM. Without any labeled data or external resource, SMTM only needs few seed words relevant to each category to conduct multi-label classification. The experimental results on two public datasets show that SMTM outperforms existing state-of-the-art dataless baselines and some supervised techniques. We further discussed some scenarios in which SMTM is a more disired choice than supervised solutions. Our approach is preferable when training data is not large or the quality of the data cannot be guaranteed. It is also disirable to use our approach only on the precisely described categories so that labeled documents are not required for those labels.

For future works, we plan to test whether dataless setting is applicable to extreme multi-label classification. Our experiments are conducted on two normal-sized multi-labeled datasets. In real applications, the number of categories could reach hundreds of thousands or millions. It is interesting to check whether SMTM can work in this situation. In our evaluation, we used LDA-based strategy to select seed words for each category. However, this strategy may not work well when the dimension of the category space is extremely high. It is also necessary to develop more advanced strategies to select seed words.

%\begin{acknowledgements}
%\end{acknowledgements}

\bibliographystyle{spbasic}      % basic style, author-year citations
\bibliography{citation.bib}   % name your BibTeX data base

\begin{thebibliography}{47}
\providecommand{\natexlab}[1]{#1}
\providecommand{\url}[1]{{#1}}
\providecommand{\urlprefix}{URL }
\expandafter\ifx\csname urlstyle\endcsname\relax
  \providecommand{\doi}[1]{DOI~\discretionary{}{}{}#1}\else
  \providecommand{\doi}{DOI~\discretionary{}{}{}\begingroup
  \urlstyle{rm}\Url}\fi
\providecommand{\eprint}[2][]{\url{#2}}

\bibitem[{Belanger and McCallum(2016)}]{belanger2016structured}
Belanger D, McCallum A (2016) Structured prediction energy networks. In:
  Proceedings of the 36th Annual International Conference on Machine Learning,
  pp 983--992

\bibitem[{Blei et~al(2003)Blei, Ng, and Jordan}]{Blei2003latent}
Blei DM, Ng AY, Jordan MI (2003) Latent dirichlet allocation. {Journal of
  Machine Learning Research} 3(Jan):993--1022

\bibitem[{Chang et~al(2008)Chang, Ratinov, Roth, and
  Srikumar}]{chang2008importance}
Chang MW, Ratinov LA, Roth D, Srikumar V (2008) Importance of semantic
  representation: Dataless classification. In: Proceedings of the the 23rd AAAI
  Conference on Artificial Intelligence, pp 830--835

\bibitem[{Chen et~al(2017)Chen, Ye, Xing, Chen, and Cambria}]{chen2017ensemble}
Chen G, Ye D, Xing Z, Chen J, Cambria E (2017) Ensemble application of
  convolutional and recurrent neural networks for multi-label text
  categorization. In: Proceedings of the 2017 International Joint Conference on
  Neural Networks, pp 2377--2383

\bibitem[{Chen et~al(2015)Chen, Xia, Jin, and Carroll}]{chen2015dataless}
Chen X, Xia Y, Jin P, Carroll J (2015) Dataless text classification with
  descriptive lda. In: Proceedings of the the 29th AAAI Conference on
  Artificial Intelligence, pp 2224--2231

\bibitem[{Chen and Liu(2014)}]{chen2014mining}
Chen Z, Liu B (2014) Mining topics in documents: standing on the shoulders of
  big data. In: Proceedings of the 20th ACM SIGKDD International Conference on
  Knowledge Discovery and Data Mining, pp 1116--1125

\bibitem[{Chen et~al(2013)Chen, Mukherjee, Liu, Hsu, Castellanos, and
  Ghosh}]{chen2013leveraging}
Chen Z, Mukherjee A, Liu B, Hsu M, Castellanos M, Ghosh R (2013) Leveraging
  multi-domain prior knowledge in topic models. In: Proceedings of the 23rd
  International Joint Conference on Artificial Intelligence, pp 2071--2077

\bibitem[{Ciss{\'e} et~al(2016)Ciss{\'e}, Al-Shedivat, and
  Bengio}]{cisse2016adios}
Ciss{\'e} M, Al-Shedivat M, Bengio S (2016) Adios: Architectures deep in output
  space. In: Proceedings of the 36th Annual International Conference on Machine
  Learning, pp 2770--2779

\bibitem[{Druck et~al(2008)Druck, Mann, and McCallum}]{druck2008learning}
Druck G, Mann G, McCallum A (2008) Learning from labeled features using
  generalized expectation criteria. In: Proceedings of the 31st Annual
  International ACM SIGIR Conference on Research and Development in Information
  Retrieval, pp 595--602

\bibitem[{Fan and Lin(2007)}]{fan2007study}
Fan RE, Lin CJ (2007) A study on threshold selection for multi-label
  classification. Department of Computer Science, National Taiwan University pp
  1--23

\bibitem[{Gabrilovich and Markovitch(2007)}]{gabrilovich2007computing}
Gabrilovich E, Markovitch S (2007) Computing semantic relatedness using
  wikipedia-based explicit semantic analysis. In: Proceedings of the 20th
  International Joint Conference on Artificial Intelligence, pp 1606--1611

\bibitem[{Ghamrawi and McCallum(2005)}]{ghamrawi2005collective}
Ghamrawi N, McCallum A (2005) Collective multi-label classification. In:
  Proceedings of the 14th ACM International Conference on Information and
  Knowledge Management, ACM, pp 195--200

\bibitem[{Griffiths and Steyvers(2004)}]{griffiths2004finding}
Griffiths TL, Steyvers M (2004) Finding scientific topics. Proceedings of the
  National Academy of Sciences 101(suppl 1):5228--5235

\bibitem[{Ishwaran and Rao(2005)}]{ishwaran2005spike}
Ishwaran H, Rao JS (2005) Spike and slab variable selection: frequentist and
  bayesian strategies. Annals of Statistics pp 730--773

\bibitem[{Ji et~al(2008)Ji, Tang, Yu, and Ye}]{ji2008extracting}
Ji S, Tang L, Yu S, Ye J (2008) Extracting shared subspace for multi-label
  classification. In: Proceedings of the 14th ACM SIGKDD International
  Conference on Knowledge Discovery and Data Mining, pp 381--389

\bibitem[{Joachims(1998)}]{joachims1998text}
Joachims T (1998) Text categorization with support vector machines: Learning
  with many relevant features. Machine Learning: ECML-98 pp 137--142

\bibitem[{Ko and Seo(2004)}]{ko2004learning}
Ko Y, Seo J (2004) Learning with unlabeled data for text categorization using
  bootstrapping and feature projection techniques. In: Proceedings of the 42nd
  Annual Meeting on Association for Computational Linguistics, p 255

\bibitem[{Kusner et~al(2015)Kusner, Sun, Kolkin, and
  Weinberger}]{kusner2015word}
Kusner M, Sun Y, Kolkin N, Weinberger K (2015) From word embeddings to document
  distances. In: Proceedings of the 35th Annual International Conference on
  Machine Learning, pp 957--966

\bibitem[{Lacoste-Julien et~al(2009)Lacoste-Julien, Sha, and
  Jordan}]{lacoste2009disclda}
Lacoste-Julien S, Sha F, Jordan MI (2009) Disclda: Discriminative learning for
  dimensionality reduction and classification. In: Proceedings of the 23rd
  Annual Conference on Neural Information Processing Systems, pp 897--904

\bibitem[{Li et~al(2016{\natexlab{a}})Li, Wang, Pavlu, and
  Aslam}]{li2016conditional}
Li C, Wang B, Pavlu V, Aslam J (2016{\natexlab{a}}) Conditional bernoulli
  mixtures for multi-label classification. In: International Conference on
  Machine Learning, pp 2482--2491

\bibitem[{Li et~al(2016{\natexlab{b}})Li, Wang, Zhang, Sun, and
  Ma}]{li2016topic}
Li C, Wang H, Zhang Z, Sun A, Ma Z (2016{\natexlab{b}}) Topic modeling for
  short texts with auxiliary word embeddings. In: Proceedings of the 39th
  International ACM SIGIR conference on Research and Development in Information
  Retrieval, pp 165--174

\bibitem[{Li et~al(2016{\natexlab{c}})Li, Xing, Sun, and Ma}]{li2016effective}
Li C, Xing J, Sun A, Ma Z (2016{\natexlab{c}}) Effective document labeling with
  very few seed words: A topic model approach. In: Proceedings of the 25th ACM
  International on Conference on Information and Knowledge Management, pp
  85--94

\bibitem[{Li and Guo(2013)}]{li2013active}
Li X, Guo Y (2013) Active learning with multi-label svm classification. In:
  Proceedings of the 23rd International Joint Conference on Artificial
  Intelligence, pp 1479--1485

\bibitem[{Lin et~al(2014)Lin, Tian, Mei, and Cheng}]{lin2014dual}
Lin T, Tian W, Mei Q, Cheng H (2014) The dual-sparse topic model: mining
  focused topics and focused terms in short text. In: Proceedings of the 23rd
  International Conference on World Wide Web, pp 539--550

\bibitem[{Liu et~al(2004)Liu, Li, Lee, and Yu}]{liu2004text}
Liu B, Li X, Lee WS, Yu PS (2004) Text classification by labeling words. In:
  Proceedings of the the 19th AAAI Conference on Artificial Intelligence, pp
  425--430

\bibitem[{Liu et~al(2017)Liu, Chang, Wu, and Yang}]{liu2017deep}
Liu J, Chang WC, Wu Y, Yang Y (2017) Deep learning for extreme multi-label text
  classification. In: Proceedings of the 40th International ACM SIGIR
  Conference on Research and Development in Information Retrieval, pp 115--124

\bibitem[{Mahmoud(2008)}]{mahmoud2008polya}
Mahmoud H (2008) P{\'o}lya urn models. CRC press

\bibitem[{Mcauliffe and Blei(2008)}]{mcauliffe2008supervised}
Mcauliffe JD, Blei DM (2008) Supervised topic models. In: Proceedings of the
  22nd Annual Conference on Neural Information Processing Systems, pp 121--128

\bibitem[{Mikolov et~al(2013)Mikolov, Sutskever, Chen, Corrado, and
  Dean}]{mikolov2013distributed}
Mikolov T, Sutskever I, Chen K, Corrado GS, Dean J (2013) Distributed
  representations of words and phrases and their compositionality. In:
  Proceedings of the 27th Annual Conference on Neural Information Processing
  Systems, pp 3111--3119

\bibitem[{Mimno et~al(2011)Mimno, Wallach, Talley, Leenders, and
  McCallum}]{mimno2011optimizing}
Mimno D, Wallach HM, Talley E, Leenders M, McCallum A (2011) Optimizing
  semantic coherence in topic models. In: Proceedings of the 2011 Conference on
  Empirical Methods in Natural Language Processing, pp 262--272

\bibitem[{Ramage et~al(2009)Ramage, Hall, Nallapati, and
  Manning}]{ramage2009labeled}
Ramage D, Hall D, Nallapati R, Manning CD (2009) Labeled lda: A supervised
  topic model for credit attribution in multi-labeled corpora. In: Proceedings
  of the 2009 Conference on Empirical Methods in Natural Language Processing,
  pp 248--256

\bibitem[{Ramage et~al(2011)Ramage, Manning, and Dumais}]{ramage2011partially}
Ramage D, Manning CD, Dumais S (2011) Partially labeled topic models for
  interpretable text mining. In: Proceedings of the 17th ACM SIGKDD
  International Conference on Knowledge Discovery and Data Mining, pp 457--465

\bibitem[{Read et~al(2011)Read, Pfahringer, Holmes, and
  Frank}]{read2011classifier}
Read J, Pfahringer B, Holmes G, Frank E (2011) Classifier chains for
  multi-label classification. Machine learning 85(3):333--359

\bibitem[{Rubin et~al(2012)Rubin, Chambers, Smyth, and
  Steyvers}]{rubin2012statistical}
Rubin TN, Chambers A, Smyth P, Steyvers M (2012) Statistical topic models for
  multi-label document classification. {Machine Learning} 88(1):157--208

\bibitem[{Soleimani and Miller(2016)}]{soleimani2016semi}
Soleimani H, Miller DJ (2016) Semi-supervised multi-label topic models for
  document classification and sentence labeling. In: Proceedings of the 25th
  ACM International on Conference on Information and Knowledge Management, pp
  105--114

\bibitem[{Song and Roth(2014)}]{song2014dataless}
Song Y, Roth D (2014) On dataless hierarchical text classification. In:
  Proceedings of the the 28th AAAI Conference on Artificial Intelligence, pp
  2224--2231

\bibitem[{Song et~al(2016)Song, Upadhyay, Peng, and Roth}]{song2016cross}
Song Y, Upadhyay S, Peng H, Roth D (2016) Cross-lingual dataless classification
  for many languages. In: Proceedings of the 25th International Joint
  Conference on Artificial Intelligence, pp 2901--2907

\bibitem[{Sun et~al(2010)Sun, Zhang, and Zhou}]{sun2010multi}
Sun YY, Zhang Y, Zhou ZH (2010) Multi-label learning with weak label. In:
  Proceedings of the the 24th AAAI Conference on Artificial Intelligence, pp
  593--598

\bibitem[{Tao et~al(2012)Tao, Li, Lau, and Wang}]{tao2012unsupervised}
Tao X, Li Y, Lau RY, Wang H (2012) Unsupervised multi-label text classification
  using a world knowledge ontology. In: Proceedings of the 2012 Pacific-Asia
  Conference on Knowledge Discovery and Data Mining, pp 480--492

\bibitem[{Tsoumakas and Katakis(2006)}]{tsoumakas2006multi}
Tsoumakas G, Katakis I (2006) Multi-label classification: An overview.
  International Journal of Data Warehousing and Mining 3(3)

\bibitem[{Tsoumakas et~al(2009)Tsoumakas, Katakis, and
  Vlahavas}]{tsoumakas2009mining}
Tsoumakas G, Katakis I, Vlahavas I (2009) Mining multi-label data. In: Data
  mining and knowledge discovery handbook, Springer, pp 667--685

\bibitem[{Wang et~al(2017)Wang, Li, Pavlu, and Aslam}]{wang2017regularizing}
Wang B, Li C, Pavlu V, Aslam J (2017) Regularizing model complexity and label
  structure for multi-label text classification. arXiv preprint arXiv:170500740

\bibitem[{Wang and Blei(2009)}]{wang2009decoupling}
Wang C, Blei DM (2009) Decoupling sparsity and smoothness in the discrete
  hierarchical dirichlet process. In: Proceedings of the 23rd Annual Conference
  on Neural Information Processing Systems, pp 1982--1989

\bibitem[{Wang et~al(2016)Wang, Chen, Fei, Liu, and Emery}]{wang2016targeted}
Wang S, Chen Z, Fei G, Liu B, Emery S (2016) Targeted topic modeling for
  focused analysis. In: Proceedings of the 22nd ACM SIGKDD International
  Conference on Knowledge Discovery and Data Mining, pp 1235--1244

\bibitem[{Yang et~al(2009)Yang, Sun, Wang, and Chen}]{yang2009effective}
Yang B, Sun JT, Wang T, Chen Z (2009) Effective multi-label active learning for
  text classification. In: Proceedings of the 15th ACM SIGKDD International
  Conference on Knowledge Discovery and Data Mining, pp 917--926

\bibitem[{Zhu et~al(2009)Zhu, Ahmed, and Xing}]{zhu2009medlda}
Zhu J, Ahmed A, Xing EP (2009) Medlda: maximum margin supervised topic models
  for regression and classification. In: Proceedings of the 26th Annual
  International Conference on Machine Learning, pp 1257--1264

\bibitem[{Zubiaga et~al(2009)Zubiaga, Garc{\'\i}a-Plaza, Fresno, and
  Mart{\'\i}nez}]{zubiaga2009content}
Zubiaga A, Garc{\'\i}a-Plaza AP, Fresno V, Mart{\'\i}nez R (2009) Content-based
  clustering for tag cloud visualization. In: Proceedings of the 2009
  International Conference on Advances in Network Analysis and Mining, pp
  316--319

\end{thebibliography}


\section*{Appendix}
\subsection*{A. Seed Words for Evaluation}
We manually label some seed words for Delicious and Ohsumed based on standard LDA model. The seed words for Delicious are listed as below:

\vspace{3mm}
\noindent
\begin{longtable}{|c|l|}
\hline
Category & Seed Words \\
\hline
politics & politics, government, political, democracy, senate \\
\hline
design & design, css, gallery, designers, designer, graphic \\
\hline
programming & programming, php, javascript, python, ruby \\
\hline
java & java, eclipse, tomcat, applet \\
\hline
reference & reference \\
\hline
internet & internet, traffic \\
\hline
computer & computer, mac, drive, desktop, screen, hardware \\
\hline
education & education, students, learning, school, teachers \\
\hline
web & web, html, ajax \\
\hline
language & language, languages, french \\
\hline
science & science, scientific, brain, scientists, researchers \\
\hline
writing & writing, fiction, tales \\
\hline
culture & culture, art, music \\
\hline
history & history, collections, historical, ancient \\
\hline
philosophy & philosophy, ethics \\
\hline
books & books, book, chapter, reading, authors, readers \\
\hline
english & english \\
\hline
religion & religion, christian, church, religious, fathers, \\
~ & testament, jesus \\
\hline
grammar & grammar, idioms, verbs, verb, sentence, clause,\\
~ & punctuation \\
\hline
style & style \\
\hline
\end{longtable}
\vspace{3mm}

And the seed words for Ohsumed are listed as below:

\vspace{3mm}
\noindent
\begin{longtable}{|c|l|}
\hline
Category & Seed Words \\
\hline
Bacterial Infections & bacterial, infections, mycoses, sepsis \\
and Mycoses & ~ \\
\hline
Virus Diseases & virus, viral, measles, herpes, influenza \\
 \hline
Parasitic Diseases & parasite, parasites, malaria, \\
~ & falciparum, leishmaniasis \\
 \hline
Neoplasms & neoplasms, neoplasm, cancer, carcinoma,\\
~ &  tumor \\
 \hline
Musculoskeletal Diseases & musculoskeletal, spine, osteomyelitis \\
 \hline
Digestive System Diseases & digestive, gastric, hepatitis, bowel, biliary \\
 \hline
Stomatognathic Diseases & stomatitis, teeth, parotid, periodontal \\
 \hline
Respiratory Tract Diseases & respiratory, lung, pneumonia, bronchial \\
 \hline
Otorhinolaryngologic Diseases & otolaryngologist, ear, hearing, otitis \\
 \hline
Nervous System Diseases & nervous, nerve, neurologic, dementia, \\
~ & neurological \\
 \hline
Eye Diseases & eye, eyes, cataract \\
 \hline
Urologic and Male & urologic, urological, genital, bladder, \\
Genital diseases  & prostate, prostatic  \\
 \hline
Female Genital Diseases & genital, pregnancy, endometrial,  \\
and pregnancy Complications & endometriosis \\
\hline
Cardiovascular Diseases & cardiovascular, ventricular, heart, cardiac, \\
~ & hypertension \\
 \hline
Hemic and Lymphatic Diseases & lymphadenopathy, anemia, sickle, \\
~ & thrombocytopenia \\
 \hline
Neonatal Diseases & neonatal, neonates, abnormalities, \\
and Abnormalities & congenital, anomalies  \\
 \hline
Skin and Connective & skin, connective, tissue, rheumatoid, \\
Tissue Diseases & psoriasis, dermal  \\
 \hline
Nutritional and & nutritional, nutrition, metabolic, glucose, \\
Metabolic Diseases & insulin, diabetes, diabetic \\
 \hline
Endocrine Diseases & endocrine, thyroid, parathyroid \\
 \hline
Immunologic Diseases & immunologic, immunodeficiency, leukemia \\
 \hline
Disorders of  & disorders, injuries, trauma, fracture \\
Environmental Origin & ~ \\
 \hline
Animal Diseases & animal animals \\
 \hline
pathological Conditions, & pathological postoperative\\
 Signs and Symptoms & ~ \\
 \hline
\end{longtable}
\vspace{3mm}

\end{document}